\documentclass[secnum,leqno]{rims-bessatsu}
\usepackage{amssymb, amsmath}
\usepackage{float}
\usepackage{graphics}
\usepackage[dvips]{graphicx}
\usepackage{eepic}
\usepackage{epic}
\newtheorem{thm}{Theorem}[section]

\newtheorem{prp}[thm]{Proposition}

\theoremstyle{definition}

\theoremstyle{remark}
\newcommand{\fract}[2]{{\textstyle\frac{#1}{#2}}}
\newcommand{\fr}[2]{{\textstyle\frac{#1}{#2}}}

\newcommand{\eps}{\varepsilon}

\newcommand{\CL}{{\cal L}}
\newcommand\ZZ{{\mathbb Z}}
\newcommand\RR{{\mathbb R}}
\newcommand\NN{{\mathbb N}}

\newcommand{\balpha}{\alpha\kern -6.7pt\alpha}
\newcommand{\bbalpha}{\alpha\kern -4.95pt\alpha}

\newcommand{\CH}{{\cal H}}

\newcommand{\CN}{{\cal N}}

\newcommand{\CQ}{Q}

\newcommand{\I}{A}

\newcommand\eq{\begin{equation}}
\newcommand\en{\end{equation}}
\newcommand\bea{\begin{eqnarray}}
\newcommand\eea{\end{eqnarray}}
\newcommand\nn{\nonumber}
\newcommand\half{{\textstyle\frac{1}{2}}}

\newcommand{\One}{{\hbox{{\rm 1{\hbox to 1.5pt{\hss\rm1}}}}}}
\renewcommand{\One}{{\mathbb 1}}
\renewcommand{\One}{{\rm 1\!\!1}}

\newcommand\ep{\epsilon}

\newcommand{\Ad}{\text{AdS}_5/\text{CFT}_4}
\newcommand{\so}{{*_{\bgammao}}}

\newcommand{\bgammax}{\bar{\gamma}_{\sf x}}
\newcommand{\bgammao}{\bar{\gamma}_{\sf o}}

\newcommand{\GammaO}{\Gamma_{\sf O}}
\newcommand{\RRe}{\text{Re}}
\newcommand{\IIm}{\text{Im}}
\newcommand{\ba}{\begin{eqnarray}}
\newcommand{\ea}{\end{eqnarray}}
\newcommand{\vep}{\varepsilon}

\newcommand{\be}{\begin{equation}}
\newcommand{\ee}{\end{equation}}

\newcommand{\AdsCft}{\text{AdS}_5/\text{CFT}_4}
\newcommand{\CCP}{{\mathbb C}{\mathbb P}}
\newcommand{\KK}{\chi}
\newcommand{\GG}{\varphi}
\newcommand{\PP}{\psi}

\newcommand{\CF}{{\cal F}}
\newcommand\stepsum[1]{{\stackrel{\mbox{\tiny{step #1}}}{\sum}}}

\title{ On the  $\text{AdS}_5/\text{CFT}_4$ TBA and its analytic properties}
\author{
\textsc{Andrea Cavagli\`a}\footnote{Centre for Mathematical Science, City University London,Northampton Square, London EC1V 0HB, UK. \newline e-mail: \texttt{Andrea.Cavaglia.1@city.ac.uk}}
,~\textsc{Davide Fioravanti}\footnote{INFN-Bologna and Dipartimento di Fisica,
Universit\`a di Bologna, Via Irnerio 46, 40126 Bologna, Italy. \endgraf e-mail: \texttt{Fioravanti@bo.infn.it}}
,~\textsc{Massimo  Mattelliano}\footnote{Dip.\ di Fisica Teorica
and INFN, Universit\`a di Torino, Via P.\
Giuria 1, 10125 Torino, Italy. \newline e-mail: \texttt{Mattelli@studenti.ph.unito.it}}
and \textsc{Roberto Tateo}\footnote{Dip.\ di Fisica Teorica
and INFN, Universit\`a di Torino,  Via P.\
Giuria 1, 10125 Torino, Italy. \endgraf e-mail: \texttt{Tateo@to.infn.it}}
}
\AuthorHead{A. Cavagli\`a,~D. Fioravanti,~ M.  Mattelliano and R. Tateo}         

\VolumeNo{x}           
\YearNo{2011}           
\PagesNo{000--000}      
\begin{document}
%

\maketitle

\begin{abstract}      
The thermodynamic Bethe Ansatz  equations arising in the context of the
$\text{AdS}_5/\text{CFT}_4$ correspondence exhibit an important difference
with respect to their analogues in relativistic integrable quantum field
theories: their solutions (the Y functions) are not meromorphic functions
of the rapidity, but live on a complicated Riemann surface with an
infinite number of branch points and therefore enjoy a new kind of {\it
extended} Y-system. In this paper we review the analytic
properties of the TBA solutions, and present new information coming from
their numerical study.
An identity allowing to simplify the equations and the numerical
implementation is presented, together with various plots which highlight
the analytic structure of the Y functions.
\end{abstract}

\section{Introduction}

In 1997 Maldacena~\cite{M} (see also  \cite{GKP-W}) conjectured a correspondence between  gravity  and  conformal gauge theory  of strong/weak type and, in particular, the duality $\AdsCft$ between the supersymmetric  string theory
of type IIB on a curved space $\text{AdS}_5\times\text{S}^5$
and   $\CN=4$   super Yang-Mills (SYM).

A turning point for the study of the correspondence $\AdsCft$  was the discovery
of integrability in the 't~Hooft  limit, both for the
string theory~\cite{BPR,KMMZ} and  the conformal field theory $\text{CFT}_4$ \cite{MZ}. The 't Hooft planar limit is defined by the scaling of the colour number $N\rightarrow \infty$ and the SYM coupling $g_{YM} \rightarrow 0$ while keeping the coupling $Ng_{YM}^2=\lambda = 4 \pi^2 g^2$ finite, so that  $g$ is proportional to the string tension\footnote{Another definition for the coupling $g$ is also widely used, so that in many works is found the relation $\lambda = 16 \pi^2 g^2$.}. In this limit only planar Feynman diagrams survive~\cite{'thooft}.

The Bethe Ansatz (BA)  method was   originally  introduced by H. Bethe~\cite{Bethe1} in 1931
to study a one-dimensional quantum system of spin  $1/2$ and
determine its eigenvalues and eigenstates.
This system is also known as the Heisenberg XXX model and, since the scattering of the possible
excitations can always be factored in terms of two-body amplitudes,  it has
the feature of being integrable.

In the $\CN=4$ SYM  context, a set of Bethe Ansatz-like equations were formulated by Beisert and Staudacher ~\cite{BS} (see also ~\cite{BES}) as a useful tool
to compute the  anomalous dimensions of SYM composite single trace operators or the energy of the quantum states of the string.
However, the equations emerging in this framework do not take into account  the wrapping effects due to the finite number of elementary operators in the trace~\cite{Sieg:2005kd,AJK}.
Therefore, the Beisert-Staudacher's equations lead to exact  results  only for  asymptotically long operators and are  the analogue of the Bethe-Yang equations for scattering models describing the quantisation of momenta
for a finite number of interacting particles living on a ring with very large circumference~\cite{INT,Be-MM}.

A first important step towards the partial  solution of the wrapping problem   was made in~\cite{Bajnok:2008bm}.
Adapting  the formulas  for the finite-size L\"uscher corrections~\cite{Luscher,KM}  to the $\Ad$ context,
Bajnok and Janik  were
able to predict   the four loop contribution to  the Konishi operator. The result was
readily  confirmed by the perturbative calculations of~\cite{FSSZ}.
The ideas proposed  in~\cite{Bajnok:2008bm}  can  be applied  to higher  perturbative orders
and twists~\cite{Beccaria:2009eq,Bajnok:2009vm}
but  technical complications increase dramatically  with the loop order  and, through this approach, exact  all-loop results are  probably out of reach.

An alternative tool to find  the finite-size corrections to the ground state energy  in 2D integrable models
was proposed many years ago by Aliosha Zamolodchikov~\cite{Zamolodchikov:1989cf}  modifying  the
thermodynamic Bethe Ansatz (TBA) whose original formulation traces back to Yang and Yang~\cite{Yang:1968rm}.
The TBA method leads to a set of coupled non-linear integral equations governing, as the  parameters of the model are changed, the ground state energy of the theory on  a cylinder  exactly and non-perturbatively.

\begin{figure}[h]
\centering
\includegraphics[width=6cm]{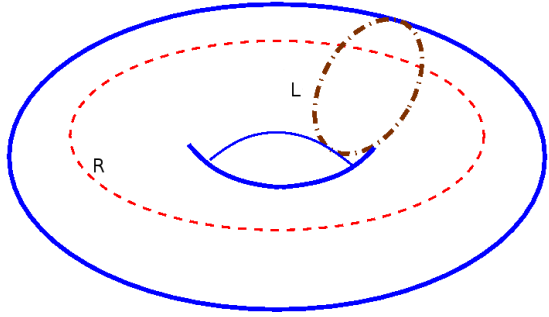}
\caption{\small The two hamiltonian descriptions of a 2D euclidean model.}
\label{fig:toro}
\end{figure}

As discussed in~\cite{Klassen:1990dx},  consider a two-dimensional euclidean  quantum field theory
defined on a torus with circumferences $R$ and $L$, as shown in Figure~\ref{fig:toro}.
The quantization of this theory is equivalently obtained by considering two alternative hamiltonian schemes.
In the \emph{direct}  scheme the system is quantised
along the direction $L$ and the hamiltonian $\CH_L$ propagates the states in the time direction $R$.
In the \emph{mirror} scheme  the system is described by the hamiltonian $\CH_R$ and the roles of $L$ and $R$ are swapped. The two descriptions  lead to the same partition function
\begin{equation}\label{funPart}
Z(R,L)=\text{Tr}(e^{-R\CH_L})=\text{Tr}(e^{-L\CH_R}).
\end{equation}
Sending $R\rightarrow\infty$, in the direct description of the theory the partition function is dominated by the smallest eigenvalue $E_0(L)$ of $\CH_L$:
\begin{equation}\label{funPart-limitDirect}
Z(R,L)=\text{Tr}(e^{-R\CH_L}) \sim e^{-RE_0(L)}.
\end{equation}
While,  in the mirror description, the $R\rightarrow\infty$  limit  corresponds
to  the thermodynamics  of a one-dimensional quantum system with hamiltonian
$\CH_R$ defined on a volume $R$ at  temperature $T = \frac{1}{L}$.
In this picture, the partition function  becomes
\begin{equation}\label{funPart-limitMirror}
Z(R,L)=\text{Tr}(e^{-L\CH_R}) \sim e^{-LRf(L)},
\end{equation}
where  $f(L)$ is the free energy per unit length at equilibrium.
Comparing  (\ref{funPart-limitDirect})  with (\ref{funPart-limitMirror})
we see that  the ground state energy of the direct theory is related to the equilibrium free energy of the mirror theory:
\begin{equation}\label{ground-state}
E_0(L) = Lf(L).
\end{equation}
Starting from the Bethe-Yang equations for the mirror theory, $f(L)$ can be found~\cite{Zamolodchikov:1989cf} adapting the method proposed  in~\cite{Yang:1968rm} for  a one-dimensional system of bosons with repulsive
delta-function interaction. As a result  of the TBA procedure, the  exact finite-size ground state energy $E_0(L)$ is   written in terms of  the  pseudoenergies $\varepsilon_a$,   solutions   to a set of coupled non-linear integral equations. In~\cite{Bazhanov:1996aq,Dorey:1996re,Dorey:1997rb,Dorey:1997yg}  the  method was extended  to excited states and an alternative but equivalent   approach was proposed in~\cite{Destri:1992qk, Fioravanti:1996rz,Feverati:1998va}
(This alternative tool has also been  used in the $\text{AdS}/\text{CFT}$ setup in~\cite{FR, Freyhult:2007pz, BFR, FFGR}).
Starting from the mirror version of the Beisert-Staudacher's equations~\cite{Arutyunov:2007tc,Arutyunov:2009zu}, the ground state TBA equations were independently  derived  in~\cite{Bombardelli:2009ns, Gromov:2009bc, Arutyunov:2009ur} and precise conjectures for particular excited state  sectors were made
in~\cite{Gromov:2009tv, Gromov:2009bc, Arutyunov:2009ax}.

Because the equations for excited states are expected to be the analytic
continuation of the ground state equations~\cite{Dorey:1996re,Dorey:1997rb}, the understanding of the analytic
structure of the TBA system is essential in the attempt to establish the results of~\cite{Gromov:2009tv, Gromov:2009bc, Arutyunov:2009ax} rigorously and to generalize them to other sectors of the theory.

The rest of this paper is organised as follows. Section~\ref{TBAM} contains
the TBA equations  of~\cite{Bombardelli:2009ns, Gromov:2009bc, Arutyunov:2009ur}. The pseudoenergies  have an infinite number of square root  branch cuts  and, basically, the $\Ad$  TBA equations
are  equivalent to a set of functional relations  containing also information on the discontinuities across the cuts~\cite{Cavaglia:2010nm}: the \emph{extended}  Y-systems described in   Section~\ref{TBAM1}.
Following~\cite{Cavaglia:2010nm}, Section~\ref{sectionDis} discusses briefly the interpretation of the TBA as dispersion relations and,
using  byproduct identities of~\cite{Cavaglia:2010nm}, a variant of the TBA equations  more suitable for  numerical integration is proposed in Section~\ref{TBAM1}.
Sections~\ref{numerical1}-\ref{numerical3} contain our preliminary numerical results for the ground state energy and a study of the pseudoenergies  in the complex rapidity plane.
Finally, Section~\ref{conclusions} contains our conclusions and  the S-matrix elements corresponding to the TBA kernels are listed in Appendix~\ref{AA}.

\section{The TBA equations}
\label{TBAM}

\begin{figure}
\begin{minipage}[h]{0.5\linewidth}
\centering
\includegraphics[width=6cm]{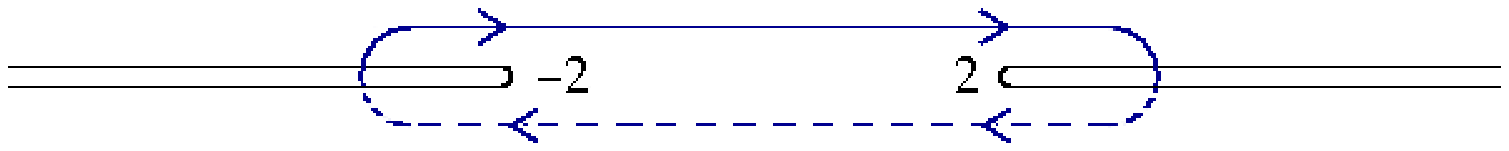}
\caption{\small The contour $\bgammao$.}
{\label{fig:gM}}
\end{minipage}
\hspace{0.3cm}
\begin{minipage}[h]{0.5\linewidth}
\centering
\includegraphics[width=6cm]{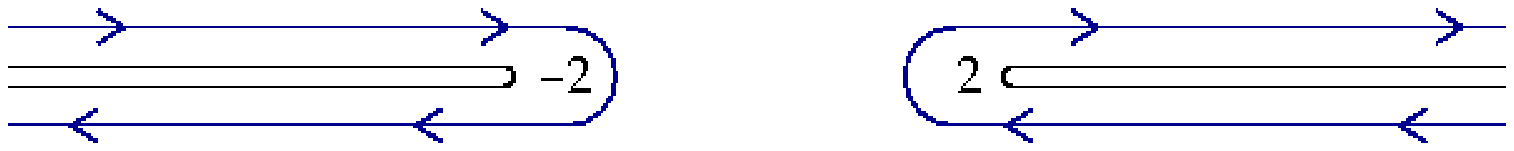}
\caption{\small The contour $\bgammax$.}
{\label{fig:gD}}
\end{minipage}
\end{figure}

The TBA equations with arbitrary chemical potentials  $\{ \mu_a \}$ are~\cite{Bombardelli:2009ns, Gromov:2009bc, Arutyunov:2009ur} :
\begin{align}
\begin{split}
\eps_{Q}(u) &= \mu_{Q} + L\,\tilde{E}_{Q}(u) - \sum_{Q' } L_{Q'}*\phi^{\Sigma}_{Q'Q}(u)\\
&+ \sum_{\alpha} {\Big (}\sum_{M} L^{(\alpha)}_{(v|M)}*\phi_{(v|M),Q}(u)
+ L^{(\alpha)}_{y}\so \; \phi_{y,Q}(u) {\Big )},\label{TBA1} \\
\end{split}
\end{align}

\begin{align}
\begin{split}
\eps^{(\alpha)}_{(v|K)}(u) &= \mu_{(v|K)}^{(\alpha)} - \sum_{Q} L_{Q}*\phi_{Q,(v|K)}(u)
+\sum_{M} L^{(\alpha)}_{(v|M)}*\phi_{M K}(u)+ L^{(\alpha)}_{y}\so \;\phi_K(u), \label{TBA2}
\end{split}
\end{align}

\begin{align}
\begin{split}
\eps^{(\alpha)}_{(w|K)}(u) &= \mu_{(w|K)}^{(\alpha)} + \sum_{M} L^{(\alpha)}_{(w|M)}*\phi_{M K}(u)
+ L^{(\alpha)}_{y}\so \; \phi_{K}(u), \label{TBA3}
\end{split}
\end{align}

\begin{align}
\begin{split}
\eps_{y}^{(\alpha)}(u) &= \mu_{y}^{(\alpha)}-\sum_{Q}  L_{Q}*\phi_{Q,y}(u)
+ \sum_{M}  (L^{(\alpha)}_{(v|M)}-L^{(\alpha)}_{(w|M)}  )*\phi_{M}(u), \label{TBA4}
\end{split}
\end{align}
where $L \in \NN $ is the  inverse of the temperature, while the sums are taken as
$\sum_{{\alpha}=1,2}$, $\sum_{K=1}^{\infty}$, with
\begin{equation*}
\tilde{E}_{Q}(u) = \ln{\frac{x(u-i\fr{Q}{g})}{x(u+i\fr{Q}{g})}}, \;\;
x(u) = \left( \frac{u}{2} -i \sqrt{ 1- \frac{u^2}{4}} \right), \;\; \IIm(x)<0,
\end{equation*}
\begin{equation} \label{La}
Y_a(u)=e^{\eps_a(u)},~~L_a(u)= \ln(1+1/Y_a(u)),
\end{equation}
and the symbols `$*$' and `$*_{\gamma}$' denote the convolutions
\begin{equation*} \label{convol}
\CF*\phi(u)=\int_{\RR} dz \; \CF(z) \,\phi(z,u),~~\CF *_{\gamma} \phi(u)=\oint_{\gamma} dz \; \CF(z) \,\phi(z,u).
\end{equation*}
The  kernels are derived from the matrix elements $S_{ab}(z,u)$ listed in Appendix~\ref{AA} using the relation: $\phi_{ab}(z,u)= \frac{1}{2 \pi i} \frac{ d}{dz} \ln S_{ab}(z,u)$.  $\phi^{\Sigma}_{Q'Q}$ in  (\ref{TBA1}) can be written as:
\begin{equation} \label{scomp}
\phi^{\Sigma}_{Q'Q}(z,u) = - \phi_{Q'Q}(z-u) - 2 K^{\Sigma}_{Q'Q}(z,u),
\end{equation}
with~\cite{Bombardelli:2009ns, Gromov:2009bc, Arutyunov:2009ur, Dorey:2007xn, Arutyunov:dressingfactor, Volin:2009uv,Volin:2010cq,Cavaglia:2010nm}
\eq
\phi_{Q'Q}(u) = \frac{1}{2 \pi i} \frac{d}{du} \ln S_{Q'Q}(u),~K_{\Gamma}^{[2]}(z-t) = \frac{1}{2 \pi i} \frac{d}{dz} \ln \frac{\Gamma(1-ig(z-t)/2)}{\Gamma(1+ig(z-t)/2)},
\en
\eq
K^{\Sigma}_{Q'Q}(z,u) =
\frac{1}{2 \pi i}\frac{d}{dz}{\ln\Sigma}_{Q'Q}(z,u) =\oint_{\bgammax} ds \; \phi_{Q', y}(z,s) \oint_{\bgammax} dt \; K_{\Gamma}^{\left[2\right]}(s- t)\phi_{y,Q}(t, u).
\label{dressingf1}
\en
The contours of integration $\bgammao$ and $\bgammax$ are represented in Figure~\ref{fig:gM} and Figure~\ref{fig:gD}, respectively.
The functions  $Y_a(u)=e^{\vep_a(u)}$ solutions of  equations  (\ref{TBA1}-\ref{TBA4}) live on
branched  coverings of the complex  u-plane with an infinite number of square root singularities
 $u = \{ \pm 2+ i m/g \}$ with $m \in \ZZ$. For the mirror $\Ad$   theory under consideration -on the
 sheet containing the physical values of the Ys-    all the cuts are
conventionally set  parallel to the real axis and external to the strip $|\RRe (u)|<2$.  This will be referred to as the reference or first  sheet.
Table~\ref{table1} shows the location of the square branch points for the various Ys.
\begin{table}[h]
\begin{center}
\begin{tabular}{|c|cc|}
\hline
Function & Singularity positions &\\
\hline
\hline
$Y^{(\alpha)}_y(u)$  & $u=\pm 2 + i\frac{2J}{g}$, & $J= 0, \pm 1, \pm 2, \dots$\\
\hline
$ Y^{(\alpha)}_{(w|M)}(u)$ & & \\
\cline{1-1}
$Y^{(\alpha)}_{(v|M)}(u)$ & $u=\pm 2 + i\frac{J}{g}$, & $J= \pm M, \pm (M+2), \pm (M+4),\dots$ \\
\cline{1-1}
$ Y_{M}(u)$ & & \\
\hline
\end{tabular}
\caption{\small Square root branch points for the  Y functions.}
\label{table1}
\end{center}
\end{table}
Finally, we shall denote with $Y^{(\alpha)}_{(y|-)}(u)$   (or simply  $Y^{(\alpha)}_y(u)$) the first sheet determination of  $Y^{(\alpha)}_y$  and with $Y^{(\alpha)}_{(y|+)}(u)=Y^{(\alpha)}_y(u_*)$ its second sheet  evaluation  obtained by analytically continuing   $u$ to  $u_*$  through the branch cut $u \in (-\infty, -2)$ (see, Figure~\ref{cutfinal}).
%
%

\section{The extended Y-system}
\label{TBAM1}

\begin{figure}
\centering
\includegraphics[width=6.5cm]{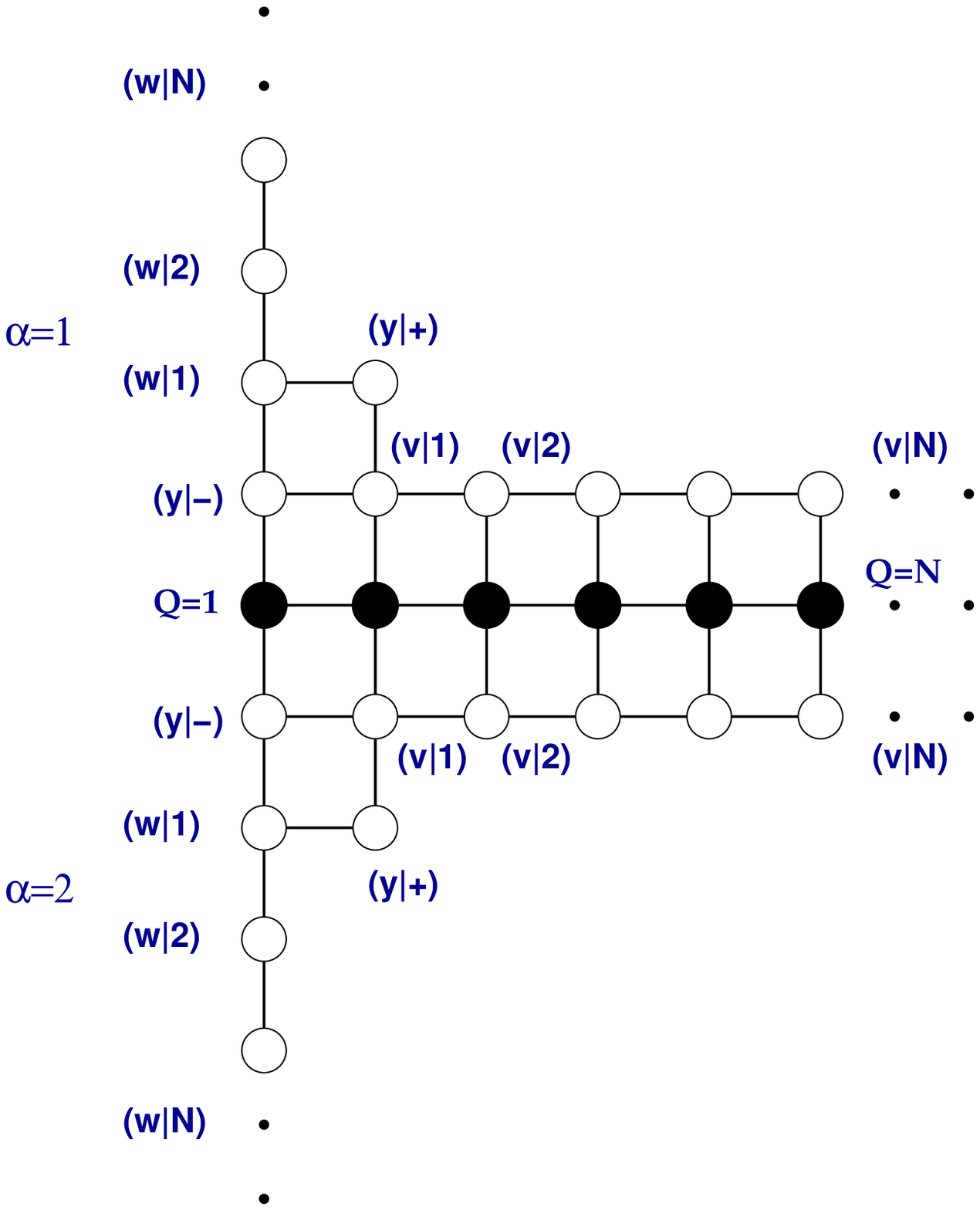}
\caption{\small The Y-system  diagram   corresponding to the  $\Ad$  TBA equations. }
\label{N4LN}
\end{figure}

The  Y-system~\cite{Zamolodchikov:1991et, Kuniba:1992ev, Ravanini:1992fi, Kuniba:2010ir} for  $\Ad$   was conjectured in~\cite{Gromov:2009tv}, rigorously  derived
in~\cite{Bombardelli:2009ns, Gromov:2009bc, Arutyunov:2009ur} and it is associated to  the
diagram represented in Figure~\ref{N4LN}. Setting
\begin{align}
\begin{split}
\Lambda_Q &=\prod_{\CQ'} e^{ \mu_{\CQ'} \; C_{\CQ' \CQ} },~~~
\Lambda^{(\alpha)}_{(y|-)} = e^{2\mu_y^{(\alpha)} - \mu_{(v|1)}^{(\alpha)}+\mu_{(w|1)}^{(\alpha)}}, \\
\Lambda^{(\alpha)}_{(w|K)} &=\prod_{M} e^{ \mu_{(w|M)}^{(\alpha)} \;C_{MK} },~~~
\Lambda^{(\alpha)}_{(v|K)} =\prod_{M} e^{\mu_{(v|M)}^{(\alpha)} \;C_{MK} },
\end{split}
\end{align}
with  $C_{MN}= 2 \delta_{M,N} - \I_{MN}$,  $\I_{1,M}=\delta_{2, M}$, $\I_{NM}= \delta_{M, N+1}+\delta_{M, N-1}$  and  $\I_{MN}=\I_{NM}$,
the Y-system with arbitrary chemical potentials is:
\eq
Y_{\CQ}(u-\fract{i}{g}) Y_{\CQ}(u+\fract{i}{g})= \Lambda_Q \prod_{\CQ'}\left(1+Y_{\CQ'}(u)\right)^{\I_{\CQ \CQ'}}
\prod_{\alpha}
\frac{ \left(1+\frac{1}{Y_{(v|\CQ-1)}^{(\alpha)}(u)}\right)^{\delta_{\CQ,1}-1}}{ {\left(1+\frac{1}{Y_{(y|-)}^{(\alpha)}(u)}\right)^{\delta_{\CQ,1}}}},
\label{Y1}
\en
\eq
Y_{(y|-)}^{(\alpha)}(u+\fract{i}{g})  Y_{(y|-)}^{(\alpha)}(u-\fract{i}{g}) =  \Lambda^{(\alpha)}_{(y|-)} \frac{\left(1+Y_{(v|1)}^{(\alpha)}(u)\right)}{\left(1+Y_{(w|1)}^{(\alpha)}(u)\right)}  \frac{1}{\left( 1+\frac{1}{Y_{1}(u)} \right)},\label{Y2}
\en
\eq
Y_{(w|M)}^{(\alpha)}(u + \fract{i}{g}) Y_{(w|M)}^{(\alpha)}(u-\fract{i}{g})= \Lambda^{(\alpha)}_{(w|M)} \prod_N\left(1+Y_{(w|N)}^{(\alpha)}(u)\right)^{\I_{MN}}
\left[\frac{\left(1+\frac{1}{Y^{(\alpha)}_{(y|-)}(u)}\right)}{\left(1+\frac{1}{Y^{(\alpha)}_{(y|+)}(u)}\right)}\right]^{\delta_{M, 1}},\label{Y3}
\en
\eq
Y_{(v|M)}^{(\alpha)}(u+\fract{i}{g}) Y_{(v|M)}^{(\alpha)}(u-\fract{i}{g})= \Lambda^{(\alpha)}_{(v|M)} \frac{\prod_N\left(1+Y_{(v|N)}^{(\alpha)}(u)\right)^{\I_{MN}}}{\left(1+\frac{1}{Y_{M+1}(u)}\right)}
\left[\frac{\left(1+Y^{(\alpha)}_{(y|-)}(u)\right)}{\left(1+Y^{(\alpha)}_{(y|+)}(u)\right)}\right]^{\delta_{M, 1}}.
\label{Y4}
\en
In relativistic  integrable models  the Y functions are in general  meromorphic  in the rapidity $u$ with zeroes and poles both linked to    $1+Y_a$ zeroes through the Y-system.
Unfortunately, the situation for  $\Ad$  is further complicated by the presence of square root branch cuts inside and at the border of the fundamental strip $|\IIm(u)|\le 1/g$.
According to the known Y $\rightarrow$ TBA transformation procedures, together with the
asymptotics of the Y functions this extra information on their analytic behaviour
should be independently supplied.

However,   the crucial discontinuity information is stored into   functions  which depend non-locally on the TBA pseudoenergies~\cite{Arutyunov:2009ur,  Cavaglia:2010nm} and  thus  on the   particular  excited state under consideration.

\begin{figure}[h]
\centering
\includegraphics[width=8cm]{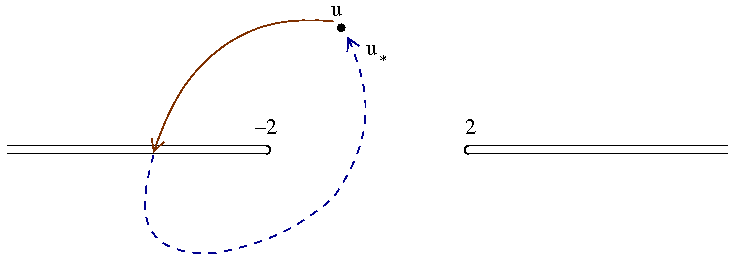}
\caption{ The second sheet image $u_*$ of $u$. }
\label{cutfinal}
\end{figure}

The main objective  of the paper~\cite{Cavaglia:2010nm} was to show how   this problem can be overcome and  that all the necessary analytic information can compactly  be encoded in the {\em basic} Y-system~(\ref{Y1}-\ref{Y4}) {\em extended}  by  the following  set  of local and  state-independent  discontinuity relations. Setting
\eq
\Delta(u)= \left[\ln Y_1(u)\right]_{+1},
\label{D0}
\en
then $\Delta$ is the function introduced in~\cite{Arutyunov:2009ur} and the  discontinuity relations are:
\eq
\left[
\Delta\right]_{\pm 2N}=\mp\sum_{\alpha=1,2} {\biggl(} \biggl[\ln\biggl(1+ \frac{1}{Y_{(y|\mp)}^{(\alpha)}}\biggr)\biggr]_{\pm 2N}+\sum_{M=1}^{N} \biggl[\ln\biggl(1+\frac{1}{ Y^{(\alpha)}_{(v|M)}}\biggr)\biggr]_{\pm (2N-M)}
+\ln{\biggl(\frac{Y_{(y|-)}^{(\alpha)}}{ Y_{(y|+)}^{(\alpha)}}\biggl)}\biggl),\label{D1}
\en
\eq
\biggl[\ln\biggl(\frac{Y^{(\alpha)}_{(y|-)}}{Y^{(\alpha)}_{(y|+)}}\biggl)\biggr]_{ \pm 2N}=- \sum_{\CQ=1}^N\left[\ln\left(1+\frac{1}{Y_{\CQ}}\right)\right]_{\pm (2N-\CQ)},
\label{D2}
\en
with $N=1,2,\dots,\infty$  and
\eq
\left[\ln{Y^{(\alpha)}_{(w|1)}} \right]_{\pm 1}= \ln\biggl( \frac{ 1+1/Y_{(y|-)}^{(\alpha)}}{  1+1/Y_{(y|+)}^{(\alpha)}}\biggl),~~~~
\left[\ln{Y^{(\alpha)}_{(v|1)}} \right]_{\pm 1}=
\ln\biggl( \frac{ 1+Y_{(y|-)}^{(\alpha)} }{  1+Y_{(y|+)}^{(\alpha)}}\biggl),
\label{D3}
\en
where the  symbol $[f]_Z$  with $Z \in \ZZ$ denotes  the discontinuity of $f(z)$
\eq
\left[ f \right]_{Z} =  \lim_{\ep \rightarrow 0^+} f(u+i Z/g+i \ep)- f(u+iZ/g -i \ep),
\label{dis0}
\en
on the  semi-infinite segments described by
$z=u+ i Z/g$ with  $u \in (-\infty,-2) \cup (2,\infty)$ and the function
$\left[ f \right(u)]_{Z}$ is the analytic extension of the discontinuity  (\ref{dis0}) to generic complex values of $u$. Finally,  $\Delta(u)$ has an additional  constant discontinuity running along the imaginary axis:
\eq
\Delta(iv+\epsilon)-\Delta(iv-\epsilon)=i2 L  \pi,~ (v\in \RR ).
\label{constantdisc0}
\en
In conclusion, while  the Y functions are defined   on an infinite sheeted Riemann surface, the {\em basic} Y-system (\ref{Y1}-\ref{Y4}) connects points lying on a single reference sheet missing a huge amount of analyticity  data.
To recover this information   it is necessary to
{\em extend} (\ref{Y1}-\ref{Y4})  by including  relations among  different Riemann sheets. This is precisely the information   contained in discontinuity relations  and, to  highlight  this fact, one could write equations (\ref{D0}-\ref{D3})  in a more  explicit   functional form involving  points from different sheets. For  example (cf (\ref{D3})):
\eq
\frac{Y^{(\alpha)}_{(w|1)}(u \pm \fract{i}{g})}{Y^{(\alpha)}_{(w|1)}(u_* \pm \fract{i}{g})}= \biggl( \frac{ 1+1/Y_{(y|-)}^{(\alpha)}(u)}{  1+1/Y_{(y|+)}^{(\alpha)}(u)}\biggl),~~~~\frac{Y^{(\alpha)}_{(v|1)}(u \pm \fract{i}{g})}{Y^{(\alpha)}_{(v|1)}(u_* \pm \fract{i}{g})}= \biggl( \frac{ 1+Y_{(y|-)}^{(\alpha)}(u)}{  1+Y_{(y|+)}^{(\alpha)}(u)}\biggl),
\label{funcdis}
\en
where $u_*$ is the second sheet image of $u$ reached by  analytic continuation through the branch cut  $u \in (-\infty,-2)$ (see, Figure~\ref{cutfinal}).
%

%
%

\section{TBA equations as  dispersion relations}
\label{sectionDis}
\begin{figure}[h]
\centering
\includegraphics[width=7cm]{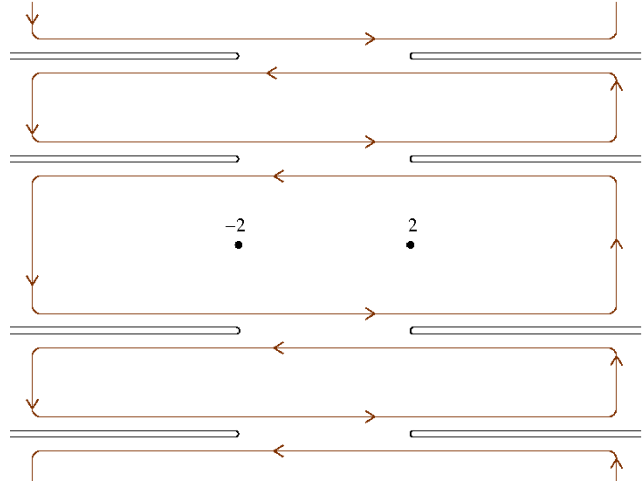}
\caption{\small  The deformed contour $\GammaO$. }
\label{Gamma0}
\end{figure}

As already mentioned in the introductory section, contrary to the more studied relativistic invariant cases, the transformation from Y-system to TBA equations is by no means  straightforward since the  local form of the    $\Ad$-related  Y-system~(\ref{Y1}-\ref{Y4}) alone does not contain information on the branch points and  it is almost totally insensitive to the precise form of  the dressing factor $\Sigma_{Q'Q}$ defined through equation (\ref{dressingf1}).
The purpose of this section  is to  give some hints on why the extension of the {\em basic}  Y-system by the discontinuity relations (\ref{D1}-\ref{D3})   resolves  this  problem completely. The interested reader is addressed to~\cite{Cavaglia:2010nm} for a more complete discussion on this important issue. Consider the following equation directly descending from (\ref{TBA4}):
\eq
\ln \left( Y^{(\alpha)}_{(y|-)}(u)/Y^{(\alpha)}_{(y|+)}(u)\right)
=-\sum_{\CQ}\int_{\RR} dz \, L_{\CQ}(z)( K(u-i\CQ/g, z)-K(u+i\CQ/g, z)  ),
\label{p1}
\en
where
\eq
K(z,u) = \frac{\sqrt{4-u^2} }{ 2\pi i \sqrt{4-z^2} } \frac{1}{ z-u}.
\label{T1}
\en
Then, it can be shown  that  the functional relation (\ref{D2}) on the  discontinuities
\eq
\left[\ln \left(Y^{(\alpha)}_{(y|-)} /Y^{(\alpha)}_{(y|+)} \right) \right]_{ \pm 2N}=- \sum_{\CQ=1}^N [L_{\CQ}]_{\pm (2N-\CQ)},
\label{dis1}
\en
combined  with some more general  analyticity information, is equivalent to (\ref{p1}).
This result comes from the observation  that the quantity:
\[
\frac{\ln \left( Y^{(\alpha)}_{(y|-)}(u)/Y^{(\alpha)}_{(y|+)}(u)\right)}{  \sqrt{4-u^2} },
\]
 is analytic at the points $u=\pm 2$, but it still has an infinite set of  branch points at $u= \pm 2 \pm  i2 N/g$ with $N \in \mathbb{N}$. From the Cauchy's integral theorem we can first write
\eq
\frac{\ln \left( Y^{(\alpha)}_{(y|-)}(u)/Y^{(\alpha)}_{(y|+)}(u)\right) }{  \sqrt{4-u^2} }
= \oint_{\gamma}  \frac{dz}{ 2 \pi i}  \frac{\ln\left( Y^{(\alpha)}_{(y|-)}(z)/Y^{(\alpha)}_{(y|+)}(z)\right)  }{(z-u)\sqrt{4-z^2} },
\label{p3}
\en
where $\gamma$ is a positive oriented contour running inside  the strip $ |\IIm(u)|< 1/g$, and then deform
$\gamma$ into the homotopically equivalent contour  $\GammaO$ represented in Figure~\ref{Gamma0} as the union of an infinite number of rectangular contours  lying between    the branch cuts of (\ref{p1}).
Restricting the analysis  to the solutions that behave asymptotically as
$\ln(Y^{(\alpha)}_{(y|-)}/Y^{(\alpha)}_{(y|+)})  \rightarrow O(1)$ uniformly as $|u| \rightarrow \infty$ the  sum of the  vertical segment contributions vanishes as the horizontal size of the rectangles tends to infinity.  Then, equation (\ref{p1}) is recovered  after inserting   relation (\ref{dis1}) and observing  that several cancelations take place between  adjacent integral contributions.
In~\cite{Cavaglia:2010nm} the entire set of TBA equations
was retrieved  starting from the \emph{extended}  Y-system and
adopting  certain minimality assumptions for the number of logarithmic
singularities in the reference Riemann sheet. These are precisely the analytic conditions fulfilled
by the  ground state TBA solutions as $\mu_y \rightarrow i \pi$.
Here, we would like to mention that is possible to show~\cite{unpublished}
that these assumptions can be relaxed, and that logarithmic singularities of the
$\eps_a$ and $L_a$ functions can lie on the first sheet without affecting the
derivation provided they are far enough from the real axis and organized in
complexes as required by the Y-system relations.
We can summarize the situation at zero chemical potentials in the following proposition.

\begin{prp}

\label{preposition}
Let the singularities on the first sheet be organized in complexes such as the
following:
\bea\label{eq:Q}
Y_{\CQ}(u_1^{(-1)})=Y_{\CQ}(u_1^{(-1)}+\frac{2i}{g})=-1;\nn\\
Y_{\CQ+1}((u_1^{(-1)}+\frac{i}{g})=Y_{\CQ-1}((u_1^{(-1)}+\frac{i}{g})=1/Y^{(\alpha)}_{(v|\CQ-1)}(u_1^{(-1)}+\frac{i}{g})=0,\nn\\
\eea
or
\bea\label{eq:v}
Y_{(v|M)}^{(\alpha)}(u_2^{(-1)})=Y_{(v|M)}^{(\alpha)}(u_2^{(-1)}+\frac{2i}{g})=-1;\nn\\
Y_{(v|M+1)}^{(\alpha)}((u_2^{(-1)}+\frac{i}{g})=Y_{(v|M-1)}^{(\alpha)}((u_2^{(-1)}+\frac{i}{g})=0;\text{
 } Y_{M+1}(u_2^{(-1)}+\frac{i}{g})=\infty^2 {\text{ ( double pole )}},\nn\\
\eea
or
\bea\label{eq:w}
Y_{(w|M)}^{(\alpha)}(u_3^{(-1)})=Y_{(w|M)}^{(\alpha)}(u_3^{(-1)}+\frac{2i}{g})=-1;\nn\\
Y_{(w|M+1)}^{(\alpha)}((u_3^{(-1)}+\frac{i}{g})=Y_{(w|M-1)}^{(\alpha)}((u_3^{(-1)}+\frac{i}{g})=0,\nn\\
\eea
or
\bea\label{eq:y}
Y_{(y|-)}^{(\alpha)}(u_4^{(-1)})=Y_{(y|-)}^{(\alpha)}(u_4^{(-1)}+\frac{2i}{g})=-1;\nn\\
Y_{(v|1)}^{(\alpha)}((u_4^{(-1)}+\frac{i}{g})=Y_{(w|1)}^{(\alpha)}((u_4^{(-1)}+\frac{i}{g})=0;\text{
 }Y_{1}(u_4^{(-1)}+\frac{i}{g})=\infty^2 {\text{ ( double pole )}},\nn\\
\eea
where, by definition, in equation (\ref{eq:Q}):
$Y_{(v|0)}^{(\alpha)}=Y_{(y|-)}^{(\alpha)}$, $Y_0=0$, and in equations (\ref{eq:v},
\ref{eq:w}): $Y_{(v|0)}^{(\alpha)}=1/Y_{(w|0)}^{(\alpha)}=Y_{(y|\pm)}^{(\alpha)}$.\\

\noindent
Moreover, let the zeroes $u^{(0)}=u^{(-1)}+\frac{i}{g}$ all lie outside the strip
$|\IIm(u)|<\frac{1}{g}$.\\
Then the \emph{extended}  Y-system implies the ground state TBA equations.\\

\noindent
The ground state equations are modified   only when the two points $u^{(-1)}$ and $u^{(-1)} + \frac{2i}{g}$
lie on different sides with respect to the real axis, which
can be interpreted as the result of a singularity having crossed the integration
contour.

\end{prp}

As we shall see in the following sections, the ground state TBA solution at  $\mu_a=0$ shows indeed several of
these complexes of singularities, in the reference sheet but far from the real axis.
%

%
%
\section{A useful variant of the  TBA equations}
\label{TBAM3}
To facilitate the numerical implementation of the equations,
the following auxiliary functions $E^{(\alpha)}$, $G^{(\alpha)}$, $T^{(\alpha)}$ and $U^{(\alpha)}$ are  introduced:
\begin{align}
\begin{split}
E^{(\alpha)}(u) &= \eps^{(\alpha)}_{(y|-)}(u)-\eps^{(\alpha)}_{(y|+)}(u) \\
&= - \sum_{Q}\int_{\RR} dz\, L_{Q}(z)\left( K(z-i\fr{Q}{g},u) - K(z+i\fr{Q}{g},u) \right),
\label{tbaE}
\end{split}
\end{align}
\begin{align}
\begin{split}
G^{(\alpha)}(u) &= \eps^{(\alpha)}_{(y|-)}(u)+\eps^{(\alpha)}_{(y|+)}(u) \\
&= 2 \sum_{M}\int_{\RR} dz\, \left( L^{(\alpha)}_{(v|M)}(z) - L^{(\alpha)}_{(w|M)}(z) \right)\,
\phi_{M}(z-u) + 2\mu^{(\alpha)}_{y} \\
&- \sum_{Q}\int_{\RR} dz\, L_{Q}(z)\,\phi_{Q}(z-u),
\label{tbaG}
\end{split}
\end{align}
\eq
T^{(\alpha)}(u) = L^{(\alpha)}_{(y|-)}(u) - L^{(\alpha)}_{(y|+)}(u),~~~~~
U^{(\alpha)}(u) = L^{(\alpha)}_{(y|-)}(u) + L^{(\alpha)}_{(y|+)}(u).
\en
Using these definitions the TBA equations (\ref{TBA1}-\ref{TBA4}) can be  recast in the   form:
\begin{align}
\begin{split}
\eps^{(\alpha)}_{(w|K)}(u) &= \mu_{(w|K)}^{(\alpha)}
+ \sum_{M}\int_{\RR} dz\, L^{(\alpha)}_{(w|M)}(z)\,\phi_{M K}(z-u)
+ \int_{-2}^{2} dz\, T^{(\alpha)}(z)\,\phi_{K}(z-u), \label{tbaW} \\
\end{split}
\end{align}
\begin{align}
\begin{split}
\eps^{(\alpha)}_{(v|K)}(u) &= \mu_{(v|K)}^{(\alpha)}
+ \sum_{M}\int_{\RR} dz\, L^{(\alpha)}_{(v|M)}(z)\,\phi_{M K}(z-u)
+ \int_{-2}^{2} dz\, ( T^{(\alpha)}(z)+E^{(\alpha)}(z) )\,\phi_{K}(z-u) \\
&- \sum_{Q}\int_{\RR} dz\, L_{Q}(z)\,\PP_{QK}(z-u),
\label{tbaV}
\end{split}
\end{align}
\begin{align}
\begin{split}
\eps_{Q}(u) &= \mu_{Q} + L\,\tilde{E}_{Q}(u)
+ \sum_{Q'}\int_{\RR} dz\, L_{Q'}(z)\,\phi_{Q'Q}(z-u) \\
&+\sum_{\alpha}\int_{-2}^{2} dz\, ( \GG^{(\alpha)}(z)+\frac{1}{2}U^{(\alpha)}(z) )
\left( K(z,u-i\fr{Q}{g}) - K(z,u+i\fr{Q}{g}) \right) \\
&- \frac{1}{2}\sum_{\alpha}\int_{-2}^{2} dz\, T^{(\alpha)}(z)\,\phi_{Q}(z-u)
+ \sum_{\alpha,M}\int_{\RR} dz\, L^{(\alpha)}_{(v|M)}(z)\,\PP_{MQ}(z-u)\\
&-\sum_{\alpha} \left( \int_{-\infty}^{-2} dz + \int_{2}^{\infty} dz \right) \, E^{(\alpha)}(z) \, \KK_{Q}(z,u),
\label{tbaQ}
\end{split}
\end{align}
with
\begin{align}
\KK_{Q}(z,u) &= \left( \int_{-\infty}^{-2} dt + \int_{2}^{\infty} dt \right) \, K_{\Gamma}^{[2]}(z-t)
 \left( K(t,u-i\fr{Q}{g}) - K(t,u+i\fr{Q}{g}) \right), \label{KKQ}
\end{align}
where the last term of (\ref{tbaQ}) is obtained from the last term of (\ref{scomp}) using the following important relation discussed  in  Appendix D of \cite{Cavaglia:2010nm}:
\begin{align}
\label{goal}
\begin{split}
\sum_{Q'} L_{Q'} \ast K_{Q'Q}^{\Sigma}(v) &=\sum_{Q'} L_{Q'} \ast \oint_{\bgammax} ds \; \phi_{Q',y}^{(\alpha)}(s)\oint_{\bgammax} dt\; K_{\Gamma}^{\left[2\right]}(s-t) \phi_{y,Q}(t,v)\\
&=-\oint_{\bgammax} ds \; \ln Y_y^{(\alpha)}(s)\oint_{\bgammax} K_{\Gamma}^{\left[2\right]}(s-t) \phi_{y,Q}(t,v).
\end{split}
\end{align}
To further optimize the implementation of the  equations, we have minimized  the number of terms that depend on the two variables  $z$ and $u$ separately~\footnote{ This was particularly useful to optimize the occupation of the computer internal  memory.}.
The last term appearing on the rhs of (\ref{tbaV}) was rewritten considering   relation (E8) in  \cite{Cavaglia:2010nm}:
\begin{align}
\begin{split}
- \sum_{Q}\int_{\RR} dz\, L_{Q}(z)\,\phi_{Q,(v|K)}(z,u) &= \int_{-2}^{2} dz\, E^{(\alpha)}(z)\,\phi_{K}(z-u) \\
&- \sum_{Q}\int_{\RR} dz\,L_{Q}(z) \psi_{Q K}(z-u),
\label{identV}
\end{split}
\end{align}
with
\begin{equation}
\PP_{Q M}(u) = \sum_{j=0}^{M-1} \phi_{Q-M+2j}(u) =
\left\{
\begin{array}{rl}
&\stepsum{2}_{j=|Q-M|}^{|Q+M|-2} \;\phi_{j}(u) \;;\; Q >M;\\ \\
&\stepsum{2}_{j=|Q-M|+2}^{|Q+M|-2} \;\phi_{j}(u) \;;\; Q \leq M.
\end{array}
\right.
\label{phiPhi}
\end{equation}
Similarly for the penultimate term on the rhs of (\ref{tbaQ}) we have used
\begin{align}
\begin{split}
\sum_{\alpha,M}\int_{\RR} dz\, L^{(\alpha)}_{(v|M)}(z)\,\phi_{(v|M),Q}(z,u) &=
\sum_{\alpha}\int_{-2}^{2} dz\, \GG^{(\alpha)}(z)\,\left( K(z,u-i\fr{Q}{g}) - K(z,u+i\fr{Q}{g}) \right) \\
&-\sum_{\alpha,M}\int_{\RR} dz\,L^{(\alpha)}_{(v|M)}(z) \psi_{MQ}(z-u),
\label{identQ}
\end{split}
\end{align}
with
\begin{align}
\GG^{(\alpha)}(u) &= \sum_{N}\int_{\RR} dz\, L^{(\alpha)}_{(v|N)}(z) \,\phi_{N}(z-u). \label{tbaGG}
\end{align}
The equations for the pseudoenergies used in the numerical   work to be described shortly correspond to an appropriately  discretized version of (\ref{tbaW}-\ref{tbaGG})  where the infinite sums are truncated to a finite number of terms $N_{max}$ and the integrals performed only in the   range $(-z_{max}; z_{max})$.
The chemical potentials are set to zero but for those associated to the  fermionic particles $(y|\pm)$ where $\mu_y=\mu^{(1)}_y=-\mu^{(2)}_y$.
The results obtained and partially discussed in this paper  correspond to  the parameters  in the ranges reported in Table~\ref{tab:setPar}.
\begin{table}[H]
\centering
\begin{tabular}{||l||c||}
\hline
Convergence threshold for  $L_a$& $\epsilon = 1\times10^{-14} $ \\
\hline
Coupling & $g\in(0,1]$ \\
\hline
Scale & $L\in(0,10]$ \\
\hline
Integration range ($-z_{max};z_{max}$) & $z_{max}=50$ \\
\hline
Number of $\Delta z$ intervals in ($-2;2$) & $N_F\in(1,150]$ \\
\hline
Truncation & $N_{max}\in[2,15]$ \\
\hline
Chemical potentials for  $(y|\pm)$ & $\mu_y\in\{(-10,10)\times[0,\pi)\}$ \\
\hline
\end{tabular}
\caption[Parameter ranges]
{\label{tab:setPar}\small  Parameter ranges.}
\end{table}
%
%
%
%

\section{ TBA numerical solution: the real axis}
\label{numerical1}
\begin{figure}[H]
\centering
\includegraphics[width=8.7cm]{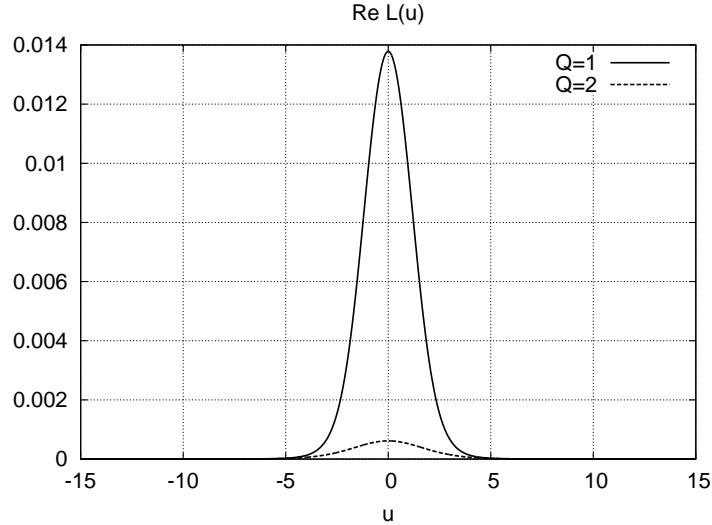}
\caption{\small Solutions $\RRe(L_{Q})$: $L=4$, $g=0.5$, $\mu_y=0$.}
\label{fig:sol-Lq}
\end{figure}

\begin{figure}[H]
\centering
\includegraphics[width=8.7cm]{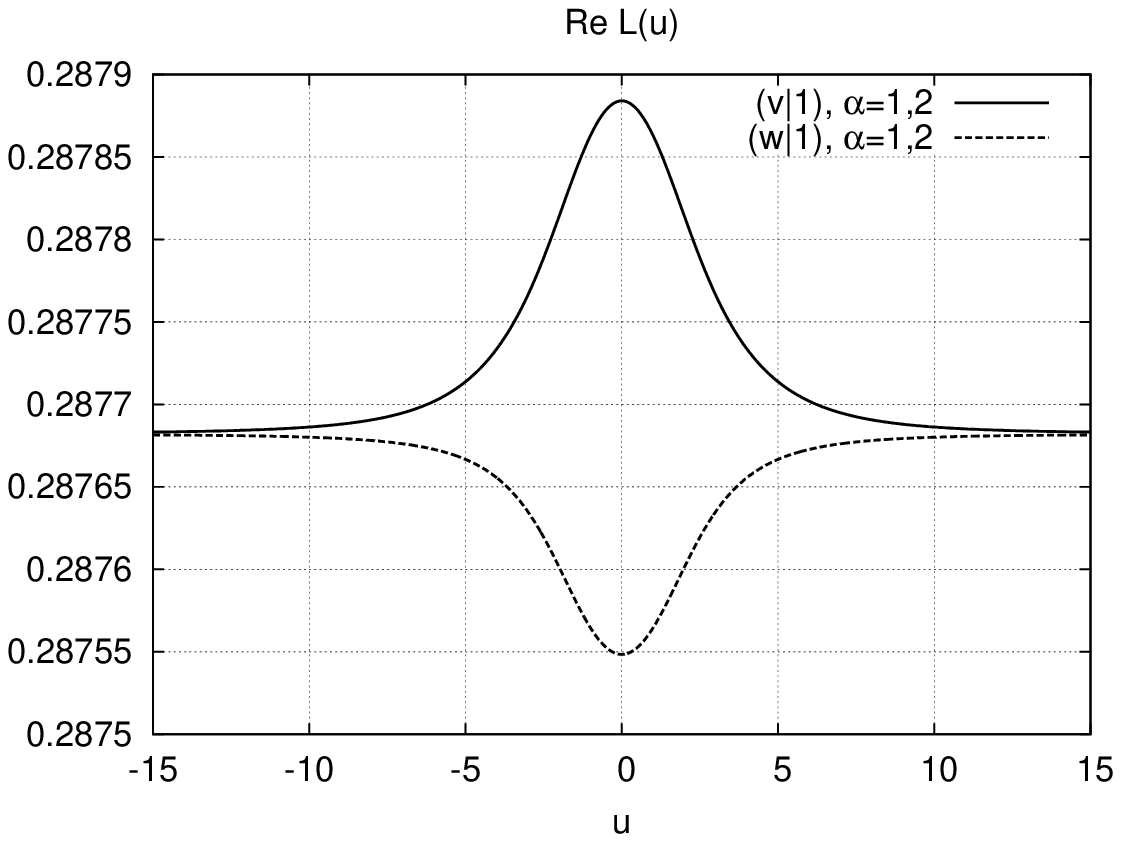}
\caption{\small Solutions $\RRe(L_{(v|1)})$, $\RRe(L_{(w|1)})$: $L=4$, $g=0.5$, $\mu_y=0$.}
\label{fig:sol-Lvw1}
\end{figure}

\begin{figure}[H]
\centering
\includegraphics[width=8.7cm]{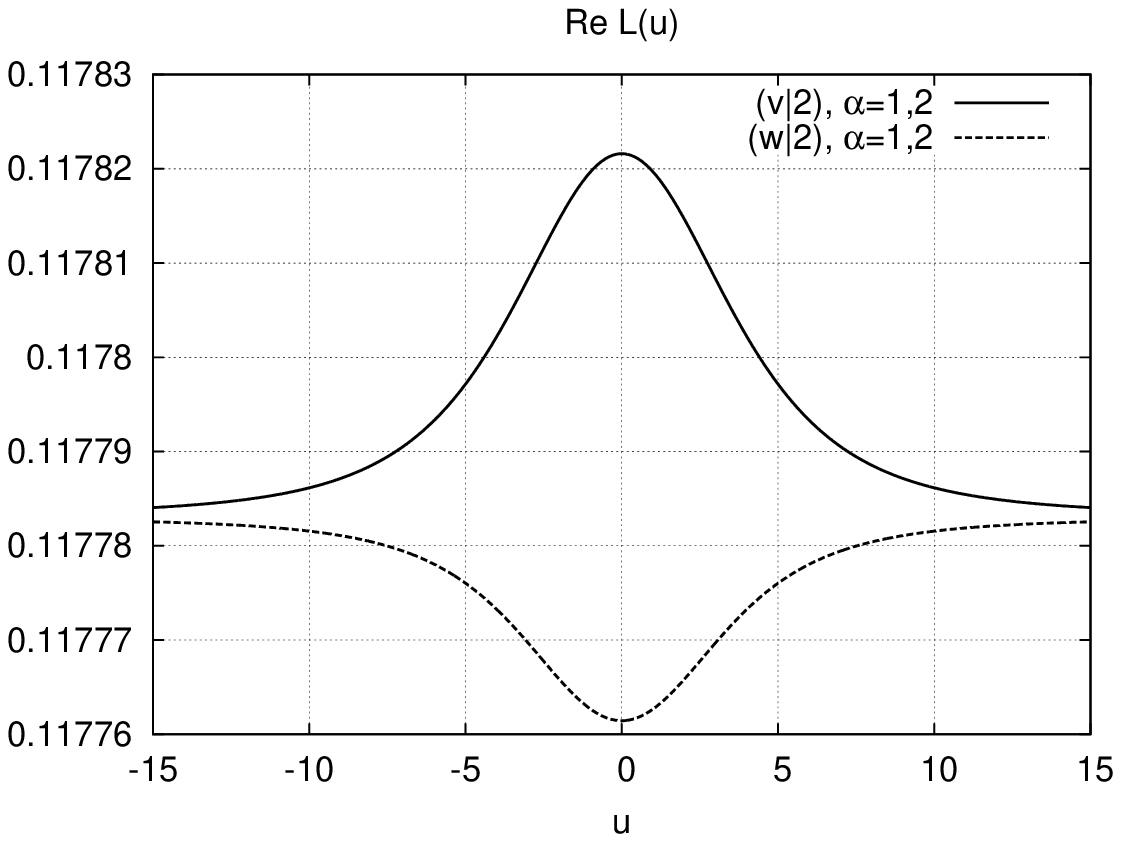}
\caption{\small Solutions $\RRe(L_{(v|2)})$, $\RRe(L_{(w|2)})$: $L=4$, $g=0.5$, $\mu_y=0$.}
\label{fig:sol-Lvw2}
\end{figure}

\begin{figure}[H]
\centering
\includegraphics[width=8.7cm]{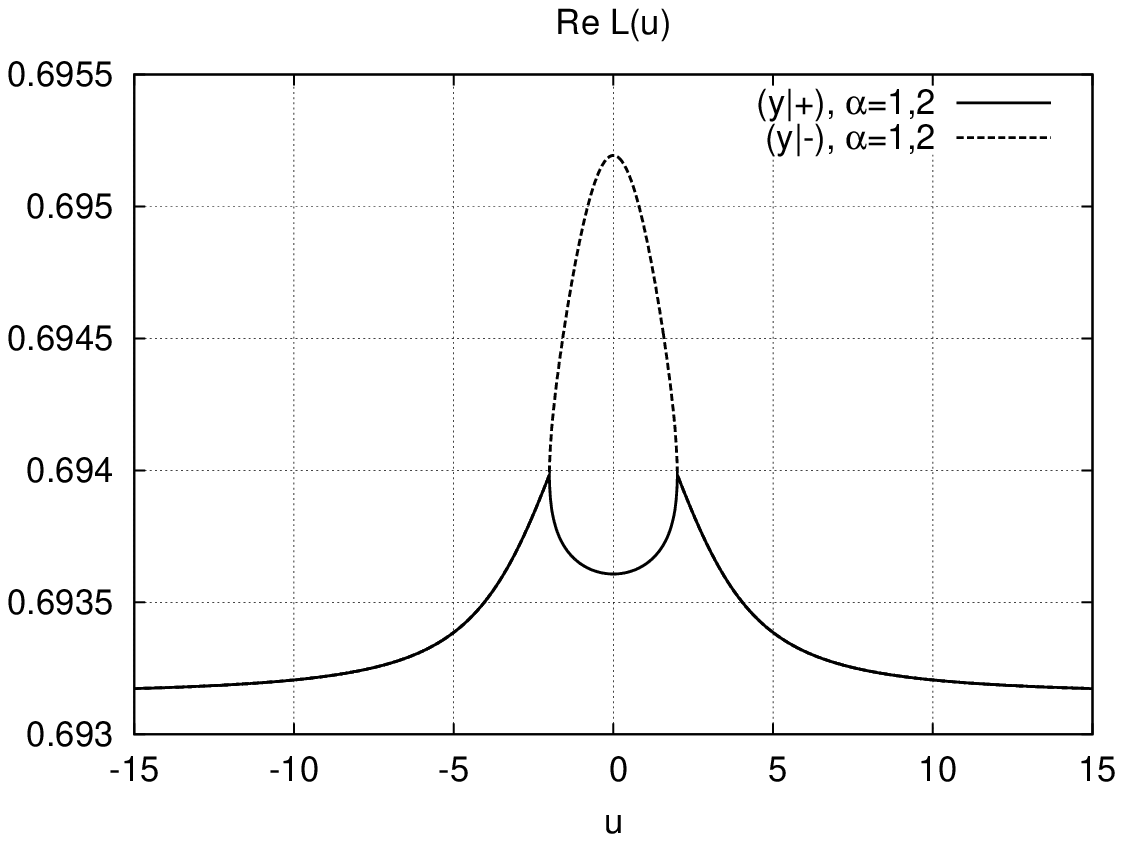}
\caption{\small Solutions $\RRe(L_{(y|\pm)})$: $L=4$, $g=0.5$, $\mu_y=0$.}
\label{fig:sol-ReLYpm}
\end{figure}

\begin{figure}[H]
\centering
\includegraphics[width=8.7cm]{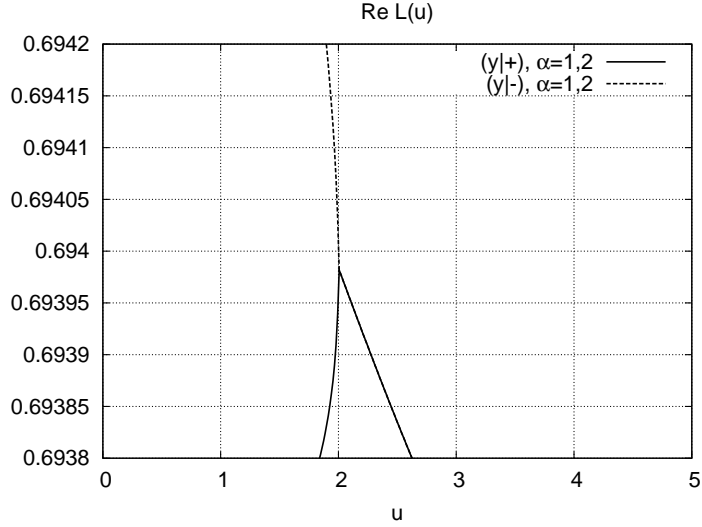}
\caption{\small The square root branch point of  $\RRe (L_{y})$: $L=4$, $g=0.5$, $\mu_y=0$.}
\label{fig:sol-ReLYmp-zoom}
\end{figure}

\begin{figure}[H]
\centering
\includegraphics[width=8.7cm]{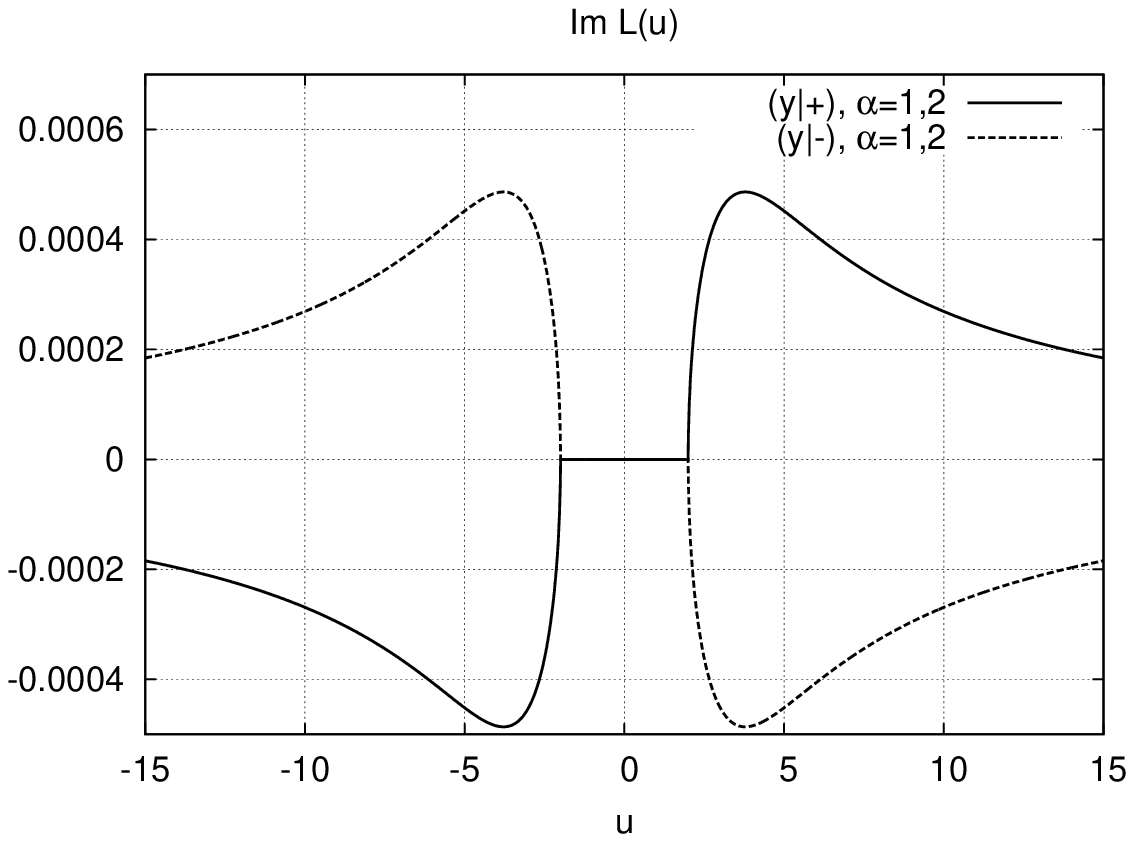}
\caption{\small Solutions $\IIm (L_{(y|\pm)})$: $L=4$, $g=0.5$, $\mu_y=0$.}
\label{fig:sol-ImLYpm}
\end{figure}

The functions $L_a(u)$, solutions  of the thermodynamic Bethe Ansatz equations,
are represented  in Figure~\ref{fig:sol-Lq}  to Figure~\ref{fig:sol-ImLYpm}.
Figure~\ref{fig:sol-Lq} shows the real parts of $L_Q(u)$ for $Q=1$ and $Q=2$ at $L=4$, $g=0.5$ and
$\mu_y=0$.
The   solutions are well localized around the origin,  tend exponentially to the asymptotic value $L_Q^0=0$
and in addition
$L_1(u) \gg  L_2(u)$.  These are   general features of the solutions of this TBA: at moderate   values of the coupling  constant
$g$  the difference between $L_a(u)$ and its asymptotic value $L_a^0$ is always  well localized
about the origin and a given  component $L_a(u)$  strongly dominates over the next, ie $L_a(u) \gg L_{a+1}(u)$.
This justifies both the relatively small  range of integration and the truncation of the sums to a finite
number of terms $N_{max}$.    Unfortunately, both the localization and the subdominancy  features   get
-for generic values of $\mu_y$ and $L$-    worse and worse as $g$ is increased above $1$ and in the strong coupling regime one has to push the computer internal resources to the extreme.
Figure~\ref{fig:sol-Lvw1} shows the results for   the excitations of type $v$ and $w$ with  $M=1$,
they also reach  the
asymptotic values $L^0_{(v | 1)}=L^0_{(w|1)}=\ln \fr{4}{3}$ exponentially fast in the large $|u|$ region.
Figure~\ref{fig:sol-Lvw2} shows  instead the real  parts of $L(u)$
for the  excitations $v,w$ with  $M=2$, they are smaller and  reach the asymptotic  value
$L^0_{(v|2)}=L^0_{(w|2)}=\ln\fr{9}{8}$ faster
that the  $M=1$ cases. Again we see that the   $M$ components dominate over the   $M+1$ ones.
The imaginary parts of $L(u)$ for the  excitations $Q$, $v$ and $w$ vanish within our computer working  precision.
Figure~\ref{fig:sol-ReLYpm} compares the real  parts for the
 $(y|\pm)$ fermionic particles. These  solutions have a physical  interpretation in terms of particle and hole densities  only in the rapidity range $u \in (-2,2)$~\cite{Bombardelli:2009ns, Gromov:2009bc, Arutyunov:2009ur}.

Indeed, one can  see cusps at $u=\pm 2$    highlighted in
Figure~\ref{fig:sol-ReLYmp-zoom}. Outside the  physical interval  $(-2,2)$ both functions tend  to the corresponding asymptotic  value $\ln 2$.

The imaginary  parts are depicted  in Figure \ref{fig:sol-ImLYpm} and they vanish for  $u \in(-2,2)$,  confirming
that only in such interval the  pseudoenergies $\eps^{(\alpha)}_{(y|\pm)}(u)$
are  physically  meaningful.
The solutions of the TBA equations expressed in terms of the functions $L_a(u) $ have,  in
principle, a maximum precision of the order of the threshold of convergence
shown in Table~\ref{tab:setPar}. In fact, the solutions are mainly affected by errors due to the
discretisation of the domain of integration and the truncation of the infinite sums.
The TBA equations are non-linear, this makes it difficult to assess a priori
the magnitude of those errors but we have  estimated the precision  by comparing solutions at various values of $N_{max}$, $z_{max}$ and $\Delta z$. As a result of this analysis, in the range of parameters reported in  Table~\ref{tab:setPar}, the estimated  precision is between $10^{-6}$  and $10^{-9}$.

%

\section{$E=0$ protected state and scale quantisation}
\label{numerical2}
The ground state energy of the model is obtained  from    the
pseudoenergies for  the  $Q$-particles as:
\begin{equation} \label{E0}
E_0(L)= -\sum_{Q=1}^{\infty}\int_{\RR} \frac{du}{2\pi} \frac{d\tilde{p}^{Q}}{du}\, L_{Q}(u),
\end{equation}
where
\begin{equation} \label{mirrorPq}
\tilde{p}^{Q}(u) = gx(u-iQ/g)-gx(u+iQ/g)+iQ,
\end{equation}
is the \emph{mirror theory}  momentum. In $\CN=4$ SYM the true
vacuum   state is protected by supersymmetry and from the TBA  this   state is
found by sending the chemical potential for the fermionic particles to~\cite{Cecotti:1992qh}
\begin{equation} \label{protectedGS}
\mu_y=\mu_{y}^{(1)}=-\mu_{y}^{(2)} \rightarrow i\pi \quad \Longrightarrow \quad E_0(L,\mu_y,g) \rightarrow  0 \quad \forall g,L,
\end{equation}
while all the other chemical potentials are kept at zero.

\begin{figure}[H]
\centering
\includegraphics[width=8.7cm]{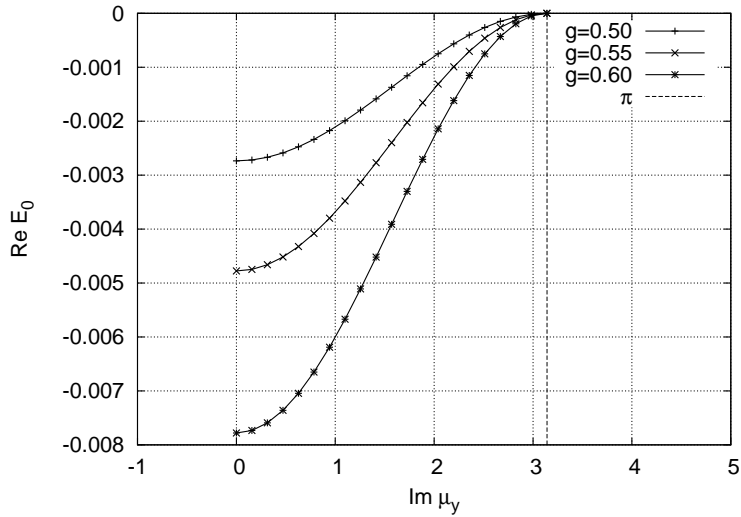}
\caption{\small $E_0$ versus $\mu_y$ and $g$: $L=4$.}
\label{fig:E0vsG-1}
\end{figure}

\begin{figure}[H]
\centering
\includegraphics[width=8.7cm]{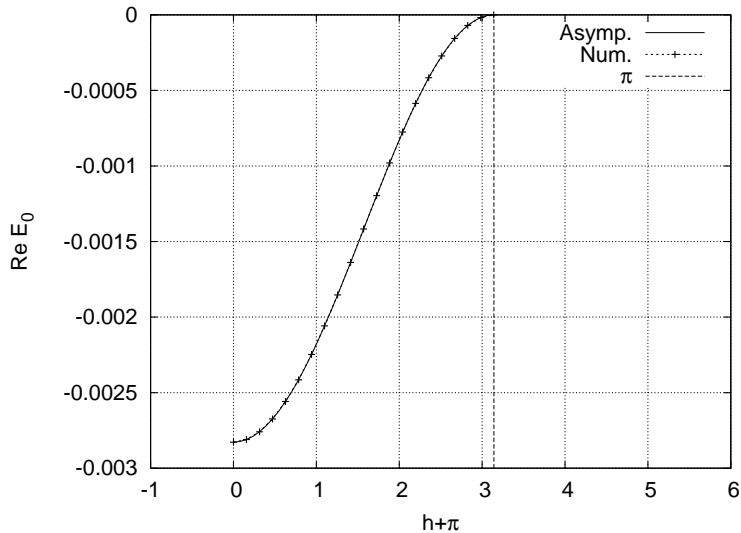}
\caption{\small $E_0$, numerical versus the exact asymptotic result: $L=4$, $g=0.5$.}
\label{fig:E0largeL}
\end{figure}

Figure~\ref{fig:E0vsG-1} shows  the occurrence of this condition at  different values of the
parameters while Figure~\ref{fig:E0largeL} compares the numerical result with the asymptotic estimate given in~\cite{Frolov:2009in}:
\begin{equation} \label{gen-h_large-L}
E_{0} \sim -\sum_{Q=1}^{\infty}\int_{\RR} \frac{du}{2\pi} \frac{d\tilde{p}^{Q}}{du}\, \ln\left(1+16Q^2 \sin^2\!\frac{h}{2}e^{-L\tilde{E}_Q(u)}\right)\;;\; \mbox{ general $h$, large $L$.}
\end{equation}
with $h = -i\mu_y-\pi$.
Further,  from the point of view of   $\CN=4$ super Yang-Mills
the parameter $L$ should be restricted to strictly positive integer values.
In fact, $L$ is  related to the number of elementary operators building the composite  trace operator of which we
would like to determine the anomalous dimension.
On the contrary, within  the TBA setup  $L$ is simply the  inverse of the temperature  and
does not  necessarily need be quantised.  In \cite{Frolov:2009in} and more generally in~\cite{Cavaglia:2010nm}    a drastic simplification  of the  analytic properties of the solutions was observed at integer $L$ but, up to now, these mathematical facts have not been neatly linked to the   gauge theory origin  of the equations.

\begin{figure}[H]
\centering
\includegraphics[width=8.7cm]{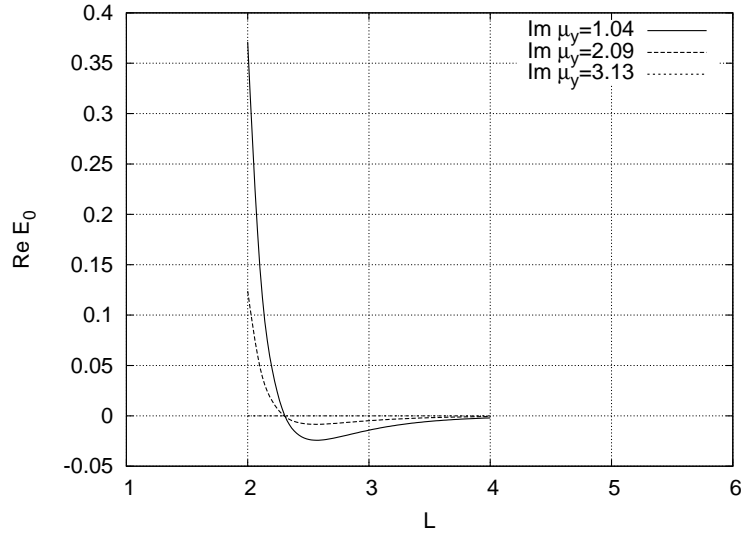}
\caption{\small  $E_0$ versus $L$, $\mu_y$: $g=0.5$, $N_{max}=2$.}
\label{fig:E0vsL-2}
\end{figure}

\begin{figure}[H]
\centering
\includegraphics[width=8.7cm]{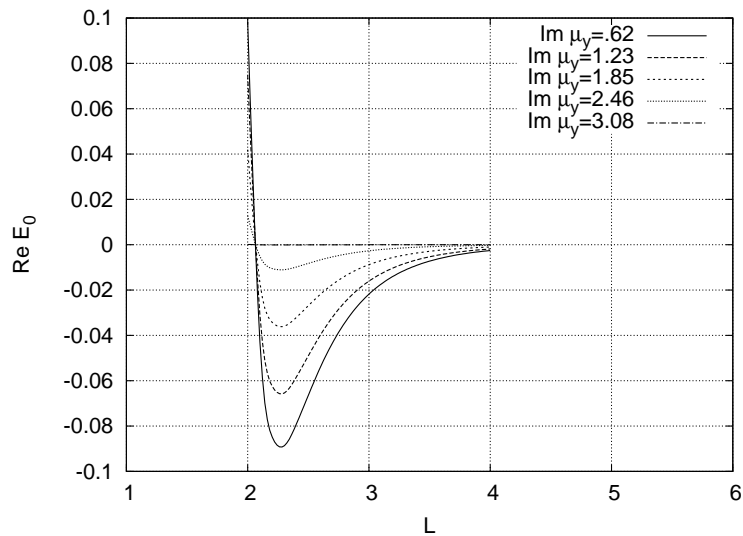}
\caption{\small  $E_0$ versus $L$, $\mu_y$: $g=0.5$, $N_{max}=5$.}
\label{fig:E0vsL-5}
\end{figure}

Indeed, the pseudoenergies  change smoothly on the real axis with the scale  and the chemical potential $\mu_y$
giving   a smooth variation of $E_0(L, \mu_y)$, as shown in
Figures~\ref{fig:E0vsL-2} and \ref{fig:E0vsL-5}.\\
Figure~\ref{fig:E0vsL-2} corresponds to a truncation at $N_{max}=2$, and  we see that the energy tends uniformly
to  zero as  $ \mu_y  \rightarrow  i \pi $.
Notice also  that, regardless of $\mu_y$,  there is a particular value of $L \approx 2.3$ where the curves simultaneously intersect  the horizontal axis.
Figure \ref{fig:E0vsL-5} corresponds to a truncation $N_{max}=5$, again the ground state
energy tends uniformly to zero as $\mu_y \rightarrow i \pi $ and at  $L \approx 2.07$
the various curves intersect on the horizontal axis.
Increasing  $N_ {max}$  we observed that the intersection  point approaches  $L=2$,
but unfortunately  the program fails to converge at $ L = 2$ for $N_ {max} \ge 8$.
%
%
\section{Mapping the complex plane}
\label{numerical3}
%
In this section we shall describe qualitatively the analytic properties of the TBA solutions
in the complex rapidity plane.
Following \cite{Dorey:1996re, Dorey:1997rb} the solutions for complex values of the rapidity were first obtained
in the fundamental strip $|\IIm(u)| \le \fr{1}{g}$
starting from the  $L_a(u)$ computed on the real axis and  using the TBA equations as integral
representations for the pseudoenergies.

\begin{figure}[h]
\centering
\includegraphics[width=8cm]{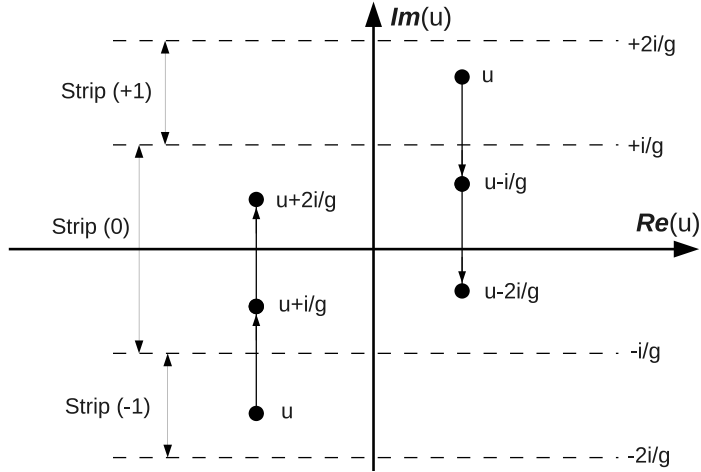}
\caption{\small  Mapping of the complex plane of rapidity $u$.}
\label{fig:cplane}
\end{figure}

\begin{figure}[H]
\centering
\includegraphics[width=14cm]{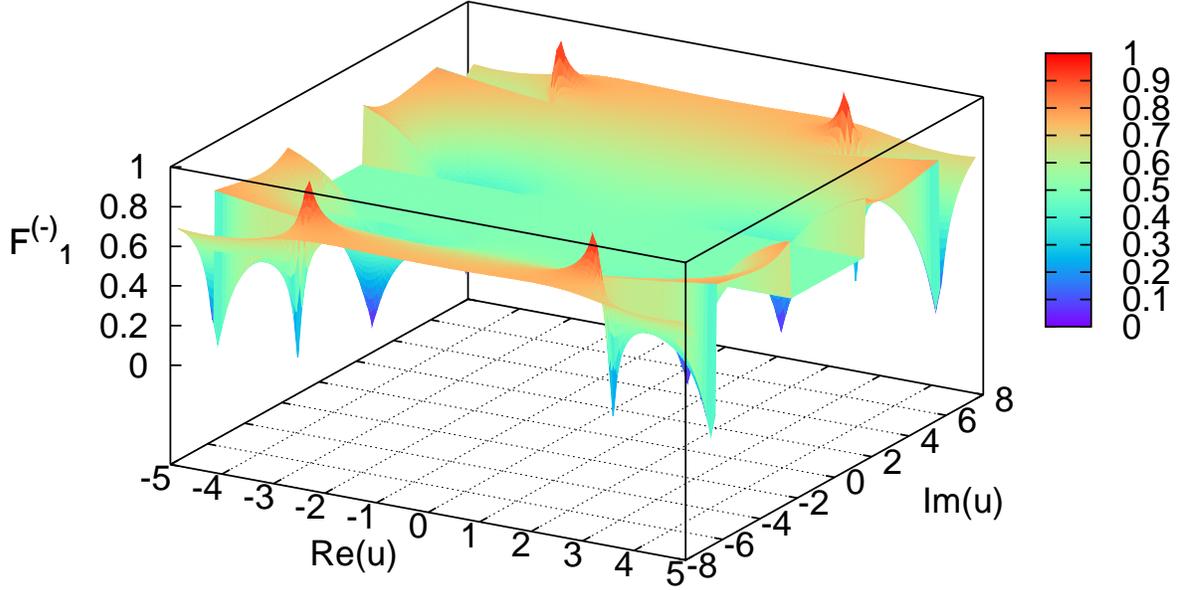}
\caption{\small  3D plot of $F^-_1(u)$: $L=4$, $g=0.5$, $\mu_y=0$.}
\label{fig:3d-FmQ1}
\end{figure}

\begin{figure}[h]
\centering
\includegraphics[width=14cm]{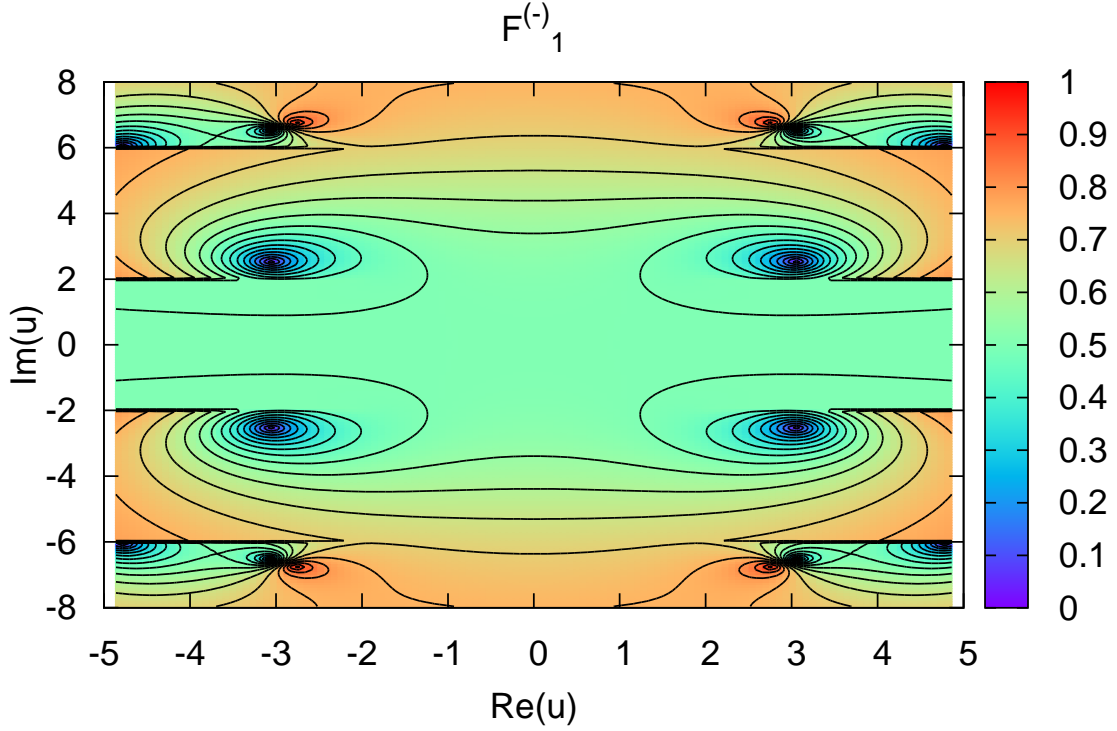}
\caption{\small  Contour plot of $F^-_1(u)$: $L=4$, $g=0.5$, $\mu_y=0$.}
\label{fig:FmQ1}
\end{figure}

Then, the solutions were calculated in the other strips, defined as
\begin{equation}\label{strip1}
\begin{split}
{\text{Strip}}^{(+k)} &= \{ u \;/ \; \RRe(u) \in (-\infty,+\infty)\;,\; \IIm(u) \in (+\fr{k}{g},+\fr{k+1}{g})\},\\
{\text{Strip}}^{(-k)} &= \{ u \;/ \; \RRe(u) \in (-\infty,+\infty)\;,\; \IIm(u) \in (-\fr{k+1}{g},-\fr{k}{g})\},
\end{split}
\end{equation}
with $k = 1,2,\cdots$, using  the {\em basic} Y-system (\ref{Y1}-\ref{Y4})
and connecting
two points in  a given strip $k$  to a
point outside it, as exemplified in Figure~\ref{fig:cplane}. In this way the whole complex plane can be mapped.
Notice that  the  {\em basic}  Y-system  does not contain a functional equation  for  $Y^{(\alpha)}_{(y|+)}$, to overcome this problem  we used   equation (\ref{tbaG})
\begin{equation}\label{Yp-G}
\eps^{(\alpha)}_{(y|+)}(u) = G^{(\alpha)}(u) - \eps^{(\alpha)}_{(y|-)}(u),
\end{equation}
with
\begin{align}\label{GinStr0}
\begin{split}
 G^{(\alpha)}(u) &= \sum_{M} \CL^{(\alpha)}_{M}*\phi_{M}(u) + 2 \mu^{(\alpha)}_y,
\end{split}
\end{align}
and $\CL^{(\alpha)}_{M}(u) = 2L^{(\alpha)}_{(v|M)}(u) - 2L^{(\alpha)}_{(w|M)}(u) - L_{M}(u)$ and analytically continued it to a generic strip in the complex plane:
\begin{equation}\label{GinStrK}
u \in {\text{Strip}}^{(\pm k)} \;;\; G^{(\alpha)}(u) = \sum_{M} \CL^{(\alpha)}_{M}*\phi_{M}(u) + 2 \mu^{(\alpha)}_y
+  \sum_{n=1}^{k}\CL^{(\alpha)}_{n}(u \mp\fr{i}{g}n).
\end{equation}
The last term on the rhs of (\ref{GinStrK}) comes  from  the  residues of the simple pole singularities
in  $\phi_{M}(u)$.
In order to  unveil  the analytical properties of the Y functions in the complex rapidity  plane  we have  introduced the functions~\cite{Dorey:1997rb}
\begin{equation}\label{Fpm}
F^{\pm}_a(u) = \frac{|1+Y^{\pm 1}_a(u)|}{1+|1+Y^{\pm 1}_a(u)|} \in [0,1],
\end{equation}
and considered the special or critical points  $u^{(0)}$,$u^{(-1)}$,$u^{(\infty)}$ such that
\begin{equation} \label{solYa}
Y_{A}(u^{(0)})=0 \;\;;\;\; Y_{A}(u^{(-1)})=-1 \;\;;\;\; Y_{A}(u^{(\infty)})\rightarrow\infty.
\end{equation}
Table \ref{tab:valCorrSol}  summarizes the values assumed by
$F^{(\pm)}_a(u)$ in correspondence  to these points. The functions
 $F^+_a$ emphasize  better the  $u^{(-1)}$ and $u^{(\infty)}$ points, on the contrary the function
$F^-_a $ are best suited to highlight the points  $u^{(0)}$ and $u^{(-1)}$.
\begin{table}[H]
\centering
\begin{tabular}{|c|c|c|c|}
\hline
$u$ & $u^{(0)}$ & $u^{(-1)}$ & $u^{(\infty)}$ \\
\hline
$Y_a(u)$ & 0 & -1 & $\infty$ \\
\hline
$F^+_a(u)$ & $\half$ & 0 & 1 \\
\hline
$F^-_a(u)$ & 1 & 0 & $\half$ \\
\hline
\end{tabular}
\caption[ $u^{(0)}$, $u^{(-1)}$, $u^{(\infty)}$.]
{\label{tab:valCorrSol}\small  $F^{\pm}$ values corresponding to $u^{(0)}$, $u^{(-1)}$, $u^{(\infty)}$.}
\end{table}
Figure \ref{fig:3d-FmQ1}  shows a three dimensional plot of  $F^-_1(u)$ where
the square root branch cuts emerge and we also  see peaks at zero ($u^{(-1)}$) and peaks at one ($u^{(0)}$). A picture of
this kind is certainly nice looking but it is not very easy to interpret.
We therefore consider two dimensional plots with contour lines, in which the value
of the functions $F^{(\pm)}_a$ is encoded using a scale of colours.
Blue corresponds to $F^{(\pm)}_a=0$ while red to $F^{(\pm)}_a=1$.
Figure~\ref{fig:FmQ1} plots  $F^-_1$,
where the critical points of type  $u^{(0)}$ and $u^{(-1)}$ and the branch cuts are easily recognisable.
The {\em basic} Y-system  relates  poles and zeroes between adjacent strips on the reference  sheet,
so that complexes or strings  of  critical points emerge. As an  example of these phenomena, we shall  highlight the link between points $u^{(0)}$, $u^{(-1)}$ and $u^{(\infty)}$ induced by  equation (\ref{Y1}) at $Q=1$:
\begin{equation} \label{Yq-bis}
Y_{1}(u-\fr{i}{g}) Y_{1}(u+\fr{i}{g}) = \left(1+Y_{2}(u)\right)
{\left(\frac{Y_{(y|-)}^{(1)}(u)}{1+Y_{(y|-)}^{(1)}(u)}\right)}^{2},
\end{equation}
where  we have used the symmetry   $Y_{(y|-)}^{(1)}=Y_{(y|-)}^{(2)}$ for $\mu^{(\alpha)}_y=0$. Equation
(\ref{Yq-bis}) shows that there are relations between the zeroes and the poles of the functions
\eq
Y_{1},\,Y_{2},\,(1+Y_{2}),\,Y^{(1)}_{(y|-)},\,(1+Y^{(1)}_{(y|-)}).
\en

\begin{figure}[H]
  \begin{center}
    \begin{tabular}{c}
      {\includegraphics[width=0.75\textwidth]{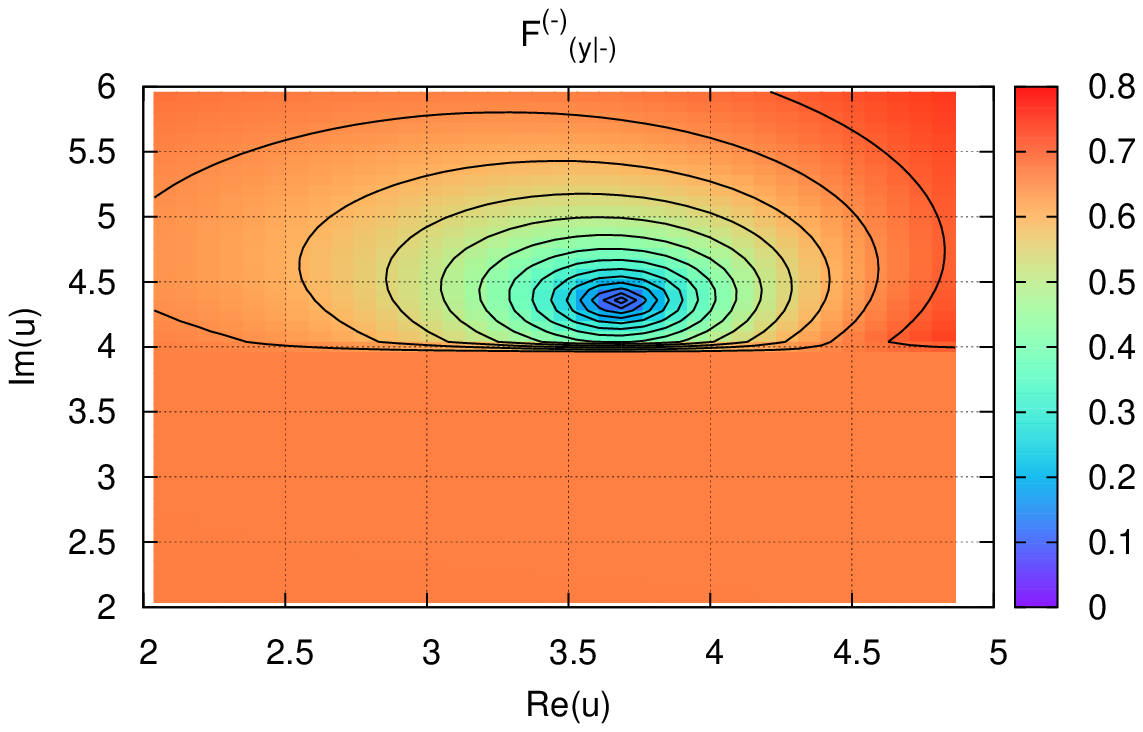}} \\
      {\includegraphics[width=0.75\textwidth]{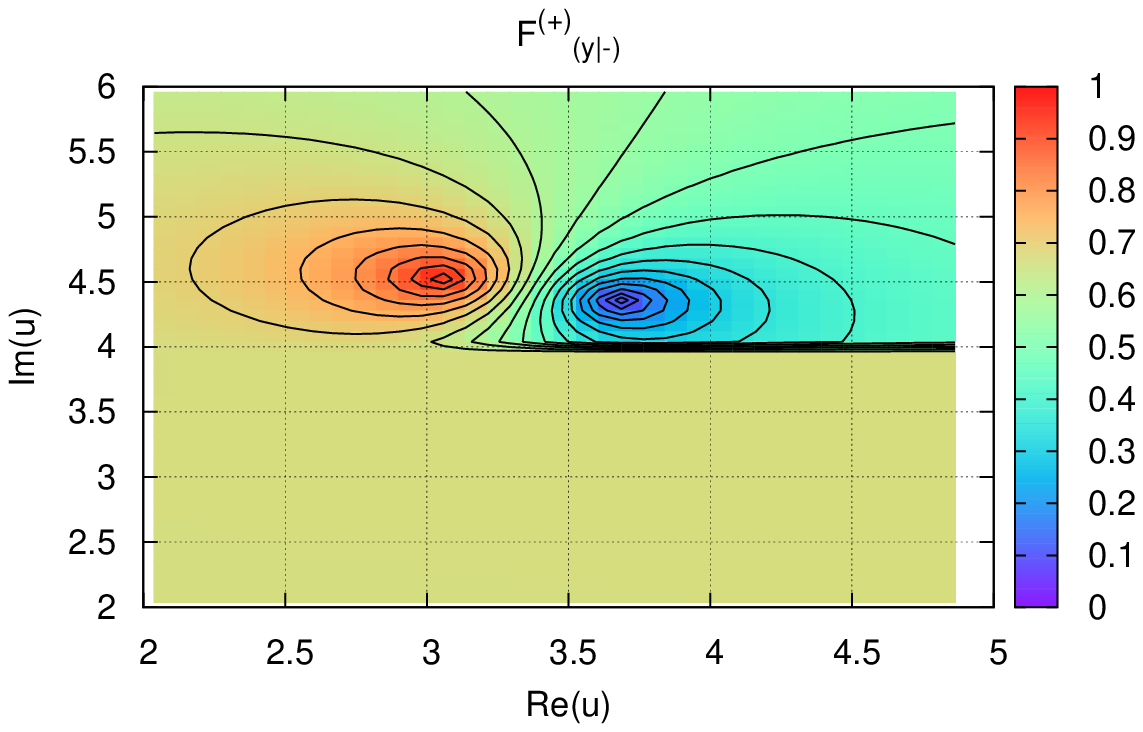}}
    \end{tabular}
    \caption[Details of $F^{\pm}_{(y|-)}(u)$: $L=4$, $g=0.5$, $\mu_y=0$.]
    {\label{fig:zoom-FpmYm}\small
    Details of $F^{\pm}_{(y|-)}(u)$: $L=4$, $g=0.5$, $\mu_y=0$.}
  \end{center}
\end{figure}

\begin{figure}[H]
  \begin{center}
    \begin{tabular}{c}
      {\includegraphics[width=0.80\textwidth]{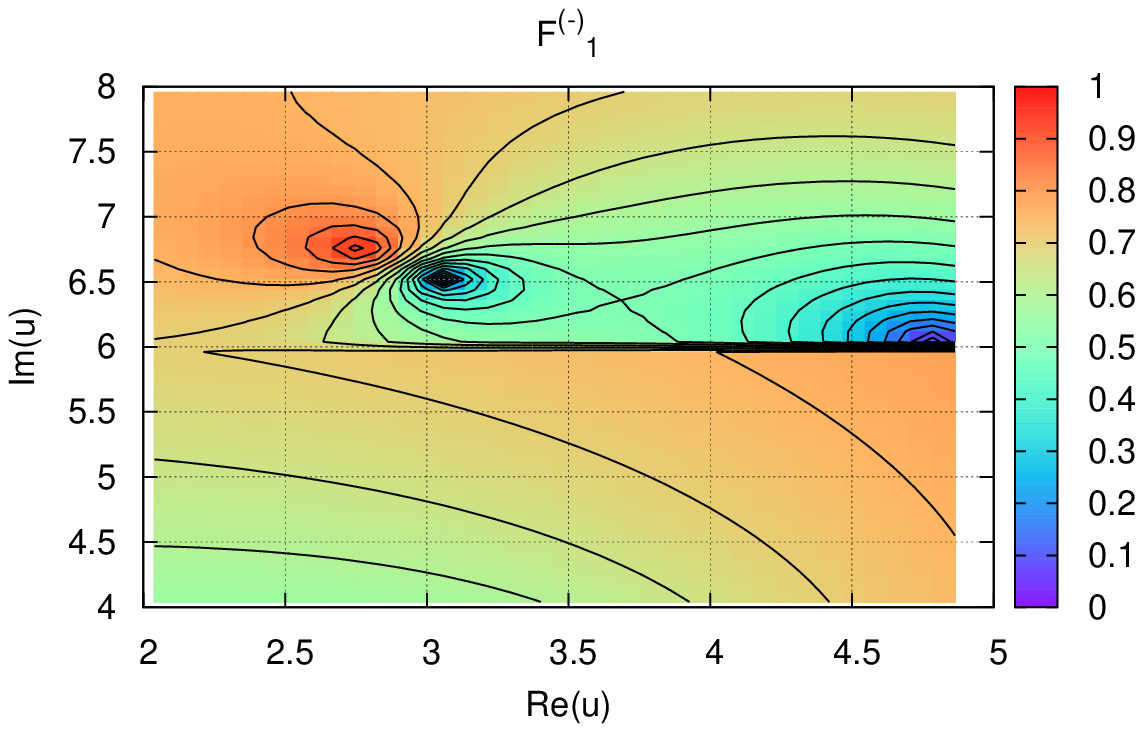}} \\
      {\includegraphics[width=0.80\textwidth]{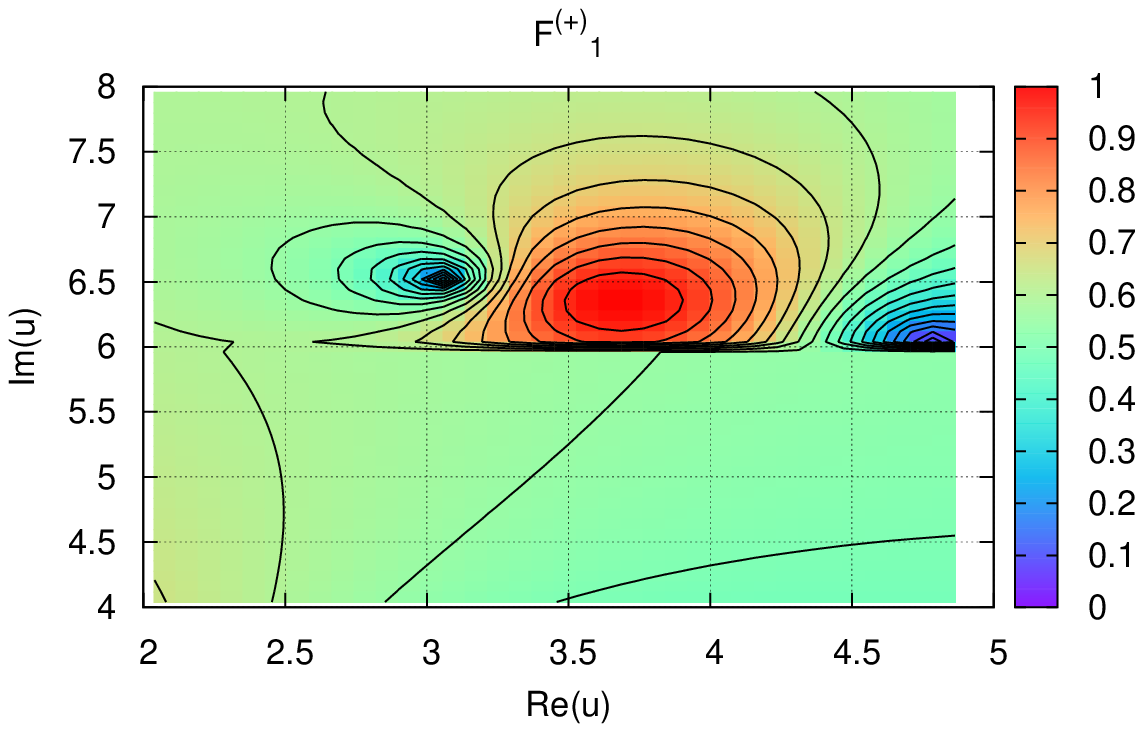}}
    \end{tabular}
    \caption[Details of $F^{\pm}_1(u)$: $L=4$, $g=0.5$, $\mu_y=0$.]
    {\label{fig:zoom-FpmQ1}\footnotesize
    Details  of $F^{\pm}_1(u)$ : $L=4$, $g=0.5$, $\mu_y=0$.}
  \end{center}
\end{figure}

\begin{figure}[H]
  \begin{center}
    \begin{tabular}{c}
      {\includegraphics[width=0.75\textwidth]{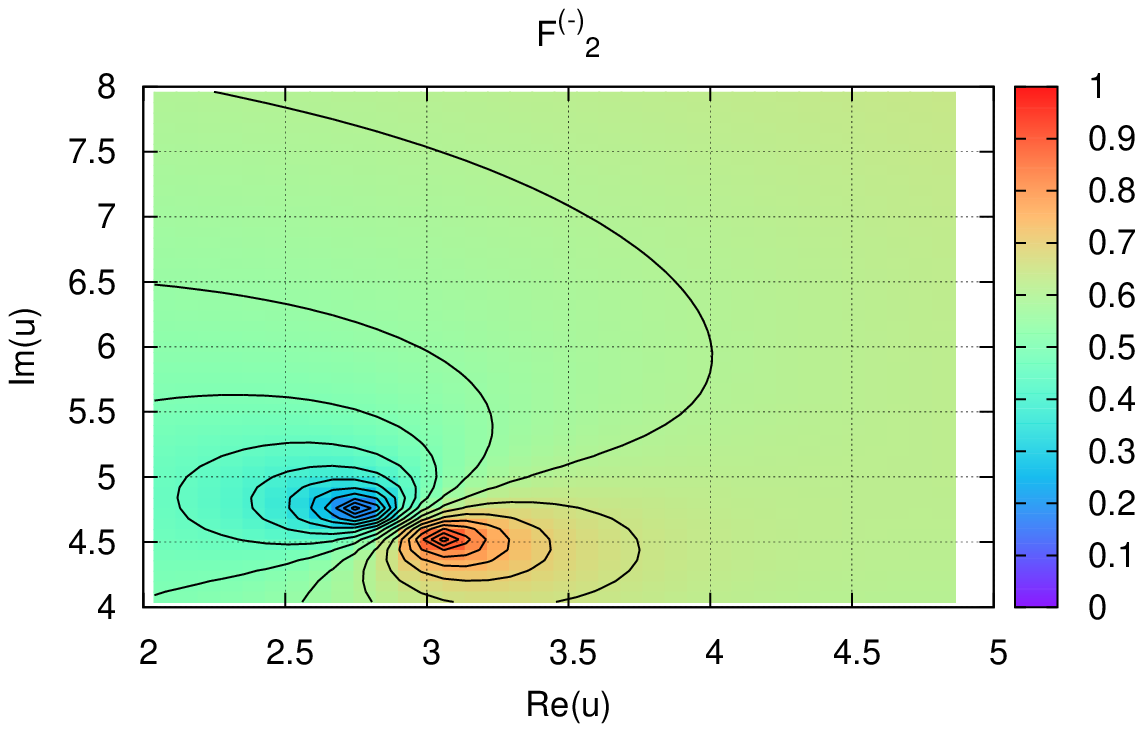}} \\
      {\includegraphics[width=0.75\textwidth]{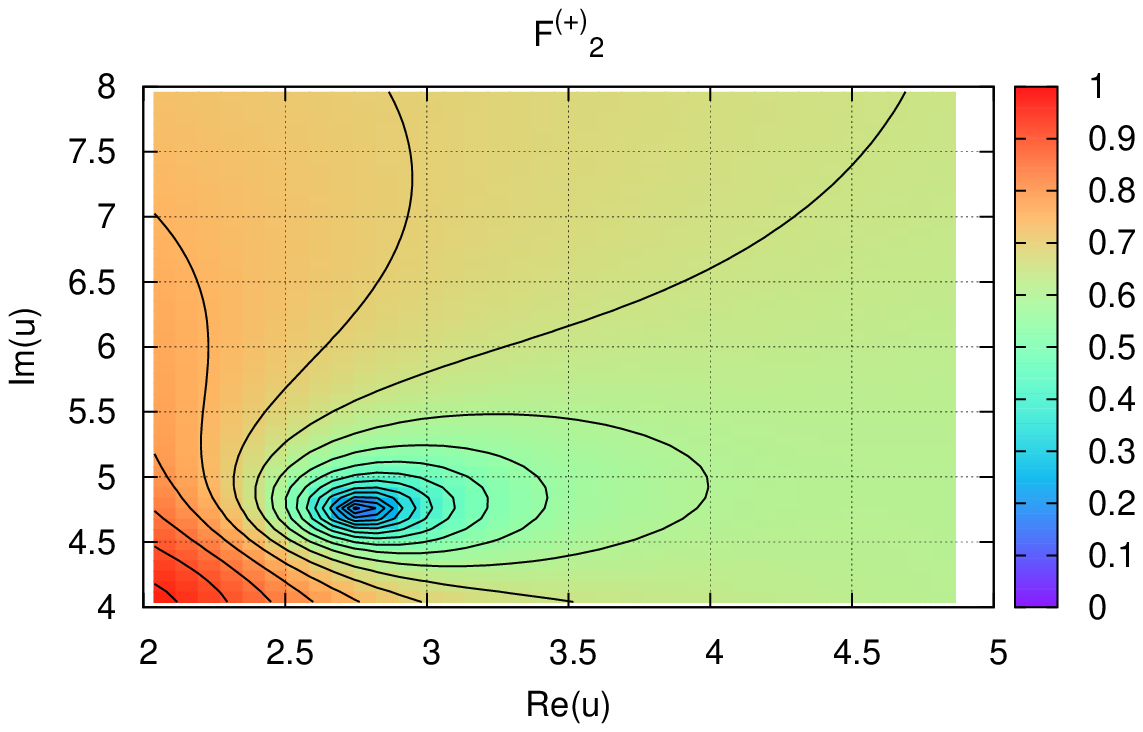}}
    \end{tabular}
    \caption[Details of $F^{\pm}_2(u)$:  $L=4$, $g=0.5$, $\mu_y=0$.]
    {\label{fig:zoom-FpmQ2}\small
    Details of $F^{\pm}_2(u)$: $L=4$, $g=0.5$, $\mu_y=0$.}
  \end{center}
\end{figure}

From  Figure~\ref{fig:zoom-FpmYm} we see that the   functions $F^{\pm}_{(y|-)}(u)$ vanish   at
$u=u^{(-1)}\approx (3.7,4.4) \in \text{Strip}^{(+2)}$ (in blue). At this point  the rhs of
equation (\ref{Yq-bis})  tends to infinity,  this implies
that  $F^+_1(u)=1$  at $u=u^{(\infty)} \approx (3.7,6.4)\in \text{Strip}^{(+3)}$ (a double pole of $Y_1$)
which can be clearly spotted in Figure~\ref{fig:zoom-FpmQ1} (in red).
Consider  now Figure~\ref{fig:zoom-FpmQ1}, the functions $F^{\pm}_{1}(u)$
show the existence of zeroes  at  $u=u^{(-1)}\approx (3.1,6.5)\in \text{Strip}^{(+3)}$ (in blue)
so at that point the  lhs  of (\ref{Yq-bis}) is equal to one. In turn, this implies
that  $F^-_2(u)=1$ at $u=u^{(0)}\approx (3.1,4.5)\in \text{Strip}^{(+2)}$,
this also clearly emerges  from   Figure~\ref{fig:zoom-FpmQ2} (in red).

This is not sufficient to guarantee compatibility between the lhs and
rhs of  (\ref{Yq-bis}), in fact also the part related to
$Y^{(1)}_{(y|-)}$ on the rhs  must tend to one. This implies that
$F^+_{(y|-)}(u)$ must have  a critical point  at $u=u^{(\infty)}\approx (3.1,4.5)\in \text{Strip}^{(+2)}$
which  presence  is clearly observable in Figure~\ref{fig:zoom-FpmYm} (in red).

Finally, the reader can  see that the function $F^+_2(u)$ plotted in Figure \ref{fig:zoom-FpmQ2}
vanishes at $u=u^{(-1)}\approx (2.8,4.7)\in \text{Strip}^{(+2)}$ (in blue),
this  is related to a critical point  $u^{(0)}\approx (2.8,6.7)\in \text{Strip}^{(+3)}$(in red)
observable in  $F^-_1(u)$ plotted in  Figure \ref{fig:zoom-FpmQ1}.
Notice that the picture emerging from this quick  inspection   matches perfectly  the  complexes of critical points defined  in Proposition~\ref{preposition}. The interest for the classification of these complexes or strings is not  purely academic  but  it is actually an important preliminary  step for the  classification of the excited states of the theory.
%
%
%

\section{Conclusions}
\label{conclusions}
The correspondence between strings and gauge theories provides   exactly soluble models  with remarkable properties.
These important systems   can be studied
using the  powerful  tools  developed over the years in the integrability  context  for the study of
models mainly relevant  to condensed matter physics in low dimensions.

A main objective  of this article  is to describe preliminary numerical results concerning a new variant of    the TBA equations for the $\Ad$ correspondence  in the attempt  to reveal the analytic properties of their solutions.

Studying the evolution of the ground state energy  at different  values  of the  scale and the chemical potential, we have checked that -as predicted by supersymmetry-  the equations correctly lead to a  vacuum state with zero energy as $\mu_y \rightarrow i \pi$ at arbitrary  scale $L$.

It was then possible to explore  the  functions $F^{(\pm)}_a$  in the complex plane of the rapidity,
allowing us  to  study  the zeroes and the singularities  of the functions $Y_a(u)$ and $1+Y_a(u)$.
We have seen  that the {\em basic} Y-system  relates  points
between adjacent strips of the complex plane and   verified the presence of
square root branch cuts exactly at the points predicted by~\cite{Cavaglia:2010nm} and summarized in Table~\ref{table1}.
These results  confirm  that the analytic properties  of the   $\Ad$ thermodynamic Bethe Ansatz equations are very different from those of the relativistic integrable quantum field theories. The Y functions  live on a complicated and -up to now- only superficially  explored  Riemann surface with an infinite number of square root branch points.

Finally our set of programs  should be  easily adaptable to other non-linear integral equations, as for example
those for  the $\text{AdS}_4/\text{CFT}_3$ correspondence, linking the
type IIA string theory on  $\text{AdS}_4\times\CCP^3$ to  the Chern-Simons model with supersymmetry
$\CN=6$  in three dimensions~\cite{Aharony:2008ug, Bombardelli:2009xz, Gromov:2009at}.

\section{Acknowledgements}
RT thanks the organizers of the conference ``Infinite Analysis 10: Developments in Quantum Integrable Systems'' and especially Atsuo Kuniba, Tomoki Nakanishi and Masato Okado  for the invitation to speak and the kind hospitality.

We acknowledge the INFN grants IS PI14, FI11, PI11 and the University PRIN 2007JHLPEZ for travel financial support.


\appendix
\section{The S-matrix elements}
\label{AA}
Here we report  the scalar factors $S_{AB}$ involved in the definition of kernels in the TBA equations (\ref{TBA1}-\ref{TBA4}).
\eq
S_{y, \CQ}(u,z)=S_{\CQ, y}(z,u) =  \left(\frac{x(z-\frac{i}{g} \CQ)-x(u)}{x(z+\frac{i}{g}\CQ)-x(u)} \right) \sqrt{\frac{x(z+\frac{i}{g}\CQ)}{x(z-\frac{i}{g}\CQ)}}.
\label{SyQ}
\en
\begin{align}
\begin{split}
S_{(v|M), \CQ}(u,z)=&S_{\CQ,(v|M)}(z,u)= \left(
\frac{x(z-\frac{i}{g}\CQ)-x(u+\frac{i}{g}M )}{x(z+\frac{i}{g}\CQ)-x(u+ \frac{i}{g} M )}\right)
\left(\frac{x(z+\frac{i}{g}\CQ)}{x(z-\frac{i}{g}\CQ)} \right) \\
&\times \left(\frac{x(z-\frac{i}{g}\CQ)-x(u-\frac{i}{g} M)}{x(z+\frac{i}{g}\CQ)-x(u-\frac{i}{g}M )}\right)\prod_{j=1}^{M-1} \left(\frac{z-u-\frac{i}{g}(\CQ-M+2j)}{z-u+\frac{i}{g}(\CQ-M+2j)}\right),
\end{split}
\end{align}
\eq
S_M(u) =  \left(\frac{u-\frac{i}{g} M }{u+\frac{i}{g} M }
\right),
\en
\begin{align}
S_{K M}(u)&= \prod_{k=1}^{K}\prod_{l=1}^{M} S_{((K+2-2k)-(M-2l))}(u)  \\
&= \left( \frac{u - \fract{i}{g} |K-M| }{ u +\frac{i}{g} |K-M|} \right) \left(
 \frac{u - \frac{i}{g} (K+M) }{ u +\frac{i}{g}(K+M)} \right)
\prod_{k=1}^{\text{min}(K,M)-1} \left( \frac{u - \frac{i}{g} (|K-M|+2k) }{ u +\frac{i}{g}(|K-M|+2k)} \right)^2
.\nn
\end{align}
The  elements $S^{\Sigma}_{\CQ'\CQ}$ are:
\eq
S^{\Sigma}_{\CQ'\CQ}(z,u)= (S_{\CQ' \CQ}(z-u))^{-1} (\Sigma_{\CQ'\CQ}(z,u))^{-2},
\en
where  $\Sigma_{\CQ'\CQ}$ is the improved dressing factor for the mirror bound states
\eq
\Sigma_{\CQ'\CQ}(z,u)=\prod_{k=1}^{\CQ'}\prod_{l=1}^{\CQ}\left(\frac{1-\frac{1}{x(z+\frac{i}{g}(\CQ'+2-2k))x(u+\frac{i}{g}(\CQ-2l))}}{1-\frac{1}{x(z+\frac{i}{g}(\CQ'-2k))
x(u+\frac{i}{g}(\CQ+2-2l))}}\right)\sigma_{\CQ' \CQ}(z,u),
\en
with $\sigma_{\CQ'\CQ}$ evaluated in the mirror kinematics. A precise analytic expression for the mirror improved dressing factor has been given in~\cite{Arutyunov:dressingfactor}, and a more compact integral representation in~\cite{Gromov:2009bc}. The equivalence between these two results was proved in \cite{Cavaglia:2010nm} .

\end{document}